\documentclass[letterpaper,twocolumn,10pt]{article}
\usepackage{usenix2019_v3}

\usepackage{tikz}
\usepackage{amsmath}
\usepackage{amsthm}
\usepackage{listings}
\usepackage{algorithm2e}
\usepackage{comment}
\usepackage{cite}
\usepackage{amsmath,amssymb,amsfonts}
\usepackage{algorithmic}
\usepackage{graphicx}
\usepackage{textcomp}
\usepackage{hyperref}
\usepackage{xcolor}
\usepackage{amsmath}
\usepackage{xcolor}
\usepackage{graphicx}
\usepackage[font=small,labelfont=bf, skip=1pt]{caption}
\usepackage{subcaption}
\usepackage{amsmath}
\usepackage{multicol}
\usepackage{listings}
\usepackage{color}
\usepackage{multirow}
\usepackage{multirow}
\usepackage{url}
\usepackage{rotating}
 \usepackage{microtype}
\usepackage{comment}
\usepackage{balance}
\usepackage{lastpage}
\usepackage{adjustbox}
\usepackage{enumitem}
\usepackage[normalem]{ulem}
\usepackage[title]{appendix}
\usepackage{authblk}

\usepackage[subtle]{savetrees}
\usepackage{url}

\begin{document}

\newcommand{\tool}[1]{\textsc{Gocc}}
\newcommand\todo[1]{\textcolor{red}{#1}}
\newcommand{\lib}[1]{\texttt{optiLib}}
\newcommand{\cs}[1]{\textbf{CS}}
\newcommand{\mutex}[1]{\texttt{Mutex}}
\newcommand{\lupair}[1]{LU-pair}
\newcommand{\fastpair}[1]{\texttt{FastLock()/FastUnlock()}}
\newcommand{\luset}[1]{\texttt{LU-Set}}
\newcommand{\milind}[1]{{\color{red} milind: #1}}
\newcommand{\adam}[1]{{\color{orange} Adam: #1}}
\newcommand{\chris}[1]{{\color{blue} chris: #1}}
\newcommand{\tim}[1]{{\color{cyan} Tim: #1}}
\newcommand{\change}[2]{\sout{#1} #2}
\newcommand{\timchange}[2]{\tim{\change{#1}{#2}}}
\newcommand{\optiLock}{\texttt{OptiLock}}
\newcommand{\optiRWLock}{\texttt{OptiRWLock}}
\newcommand{\gwt}{\texttt{GWT}}
\newcommand{\lp}{lock-point}
\newcommand{\up}{unlock-point}
\newcommand{\lup}{LU-points}
\newcommand{\speedup}{10$\times$}
\newcommand{\Eqnum}[1]{(\ref{eqn:#1})}
\newcommand{\Eqn}[1]{Equation~\Eqnum{#1}}
\newtheorem{Definition}{Definition}[section]
\newtheorem{Theorem}{Theorem}
\newtheorem{Remark}{Remark}
\newcommand{\QED}{\begin{flushright}$\Box$\end{flushright}}
\newtheorem{Observation}{Observation}
\newtheorem{Rule}{Rule}
\newtheorem{Fact}{Fact}
\newtheorem{Lemma}{Lemma}
\newtheorem{Corollary}{Corollary}
\newtheorem{Lcorol}{Corollary}
\newtheorem{Example}{Example}
\newenvironment{Proof}{\noindent{\bf Proof: }}{\qed}

\date{}

\title{\Large \bf Optimistic Concurrency Control for Real-world Go Programs \\ (Extended Version with Appendix)}

 \author[1]{Zhizhou Zhang}
 \author[2]{Milind Chabbi}
 \author[2]{Adam Welc}
 \author[1]{Timothy Sherwood}

 \affil[1]{University of California, Santa Barbara}
 \affil[1]{\textit {\{zhizhouzhang, sherwood\}@cs.ucsb.edu}}
 \affil[2]{Programming Systems Group, Uber Technologies}
 \affil[2]{\textit {\{milind,adam.welc\}@uber.com}}

\maketitle

\begin{abstract}
We present a source-to-source transformation framework, \tool{}, that consumes lock-based pessimistic concurrency programs in the Go language and transforms them into optimistic concurrency programs that use Hardware Transactional Memory (HTM). 
The choice of the Go language is motivated by the fact that concurrency is a first-class citizen in Go, and it is widely used in Go programs.
\tool{} performs rich inter-procedural program analysis to detect and filter lock-protected regions and performs AST-level code transformation of the surrounding locks when profitable.
Profitability is driven by both static analyses of critical sections and dynamic analysis via execution profiles.
A custom HTM library, using perceptron, learns concurrency behavior and dynamically decides whether to use HTM in the rewritten lock/unlock points. 
Given the rich history of transactional memory research but its lack of adoption in any industrial setting, we believe this workflow, which ultimately produces source-code patches, is more apt for industry-scale adoption.
Results on widely adopted Go libraries and applications demonstrate significant (up to $10\times$) and scalable performance gains resulting from our automated transformation while avoiding major performance regressions.
\end{abstract}

\section{Introduction}
\label{sec:intro}
Golang~\cite{effectivego} (or simply Go) is a modern programming language that has gained significant popularity over the last decade. It is being used to write enterprise software~\cite{nixwww} (e.g., to implement backend services) in some of the largest technology companies as well as to develop large and widely used open-source applications (e.g., Kubernetes~\cite{kubernetes}) and libraries (e.g., Tally~\cite{tally}). The design of Go is inspired by C, but unlike C, it supports concurrency as the first-class language construct. Even more importantly, and unlike other popular languages with first-class concurrency support (e.g., Java), the Go language goes to great lengths to simplify concurrent programming by making concurrency easy to use (and thus frequently used) by the developers~\cite{gobugs} --- any function in Go can be scheduled to execute concurrently with the rest of the code as a \emph{goroutine}~\cite{effectivego} by simply prefixing its call with the \texttt{go} keyword. %

Although Go makes writing concurrent programs easier, it still requires programmers to manage interactions between concurrently executing code --- this can be accomplished either via passing messages through channels~\cite{effectivego} or explicitly synchronizing accesses to shared memory. 
Shared memory is used more often than message passing by Go developers, and mutual exclusion via locks~\cite{gosyncPkg} remains the most widely-used synchronization mechanism across several applications~\cite{gobugs}. It is, therefore, the focus of our work.

Locks may unnecessarily serialize concurrent execution, even if the code operates on disjoint data.
Our work aims to improve the  performance of concurrent Go code, particularly code hiding behind needlessly held locks.
Our goal is to accomplish this while retaining the correctness of concurrent execution.
We utilize the concept of \emph{transactional memory} (TM)~\cite{herlihy1993transactional} to achieve this goal.
The general idea behind TM is to decide on whether two (or more) pieces of code can be executed concurrently based on whether their accesses to the underlying data are \emph{conflicting}~\cite{HerlihyBook} or not, that is, if at least one of the accesses is a write. Conflict-free executions are allowed to proceed in parallel. On the other hand, upon encountering a data access conflict, execution effects of at least one piece of code have to be \emph{rolled back} (i.e., undone), and the computation must be restarted. 
TM machinery, which originally started in software (STM)~\cite{stm-tires, tl2, ring-stm, mcrt-stm}, is now available in commodity hardware as Hardware Transactional Memory (HTM)~\cite{htm-early, intel-tm, htm-blue}. However, despite almost three decades of work in this area, TM's promise of accelerating concurrent computations for real-life software has not been quite fulfilled. We speculate that there are two reasons why this is the case.

The first reason is that TM, while being a single concept, may have different realizations in terms of algorithms and implementations (e.g., eager vs. lazy versioning~\cite{stm-isolation}) and different integration strategies at the language level (e.g., API-level solutions~\cite{mcrt-stm} or the compiler-assisted \texttt{atomic} construct used to demarcate TM-managed concurrent code~\cite{tm-lang-support}), resulting in different behavior from the programmer's perspective. Consequently, attempts to introduce TM as a separate language-level mechanism lead to significant semantic dissonance with respect to existing concurrency-related mechanisms~\cite{c-stm-semantics, java-stm-semantics}.%

The second reason is that a lot of TM (particularly STM) work was focusing on designing and implementing TM algorithms but limiting empirical evaluation to synthetic benchmarks (e.g., STMBench7~\cite{stmbench}) or measuring the performance of only selected concurrent data structures. 
Unfortunately, unlike what was expected, TM techniques did not easily generalize to real-life applications~\cite{stm-tires}. 
A few attempts to apply TM to production code were unsuccessful (e.g., an attempt to rewrite the Quake game server to use TM~\cite{tm-quake}).

In this work, we attempt to rectify some of these limitations and show that TM can be effective in accelerating real-life concurrent code. 
We focus less on the algorithmic side of TM (we use state-of-the-art off-the-shelf HTM implementation from Intel), and more on how and when to apply the TM machinery to maximize the benefit. Additionally, we replace Go locks with HTM constructs without changing the code's behavior in any way, which allows us to completely bypass complications related to transactional memory semantics.
More specifically, we employ \emph{transactional lock elision} (TLE)~\cite{Rajwar2001} --- a well-known technique that attempts to execute a lock-protected critical section as an atomic hardware transaction, reverting to using the lock if these attempts fail. 

\begin{figure}[!t]
    \centering
    \includegraphics[width=\linewidth]{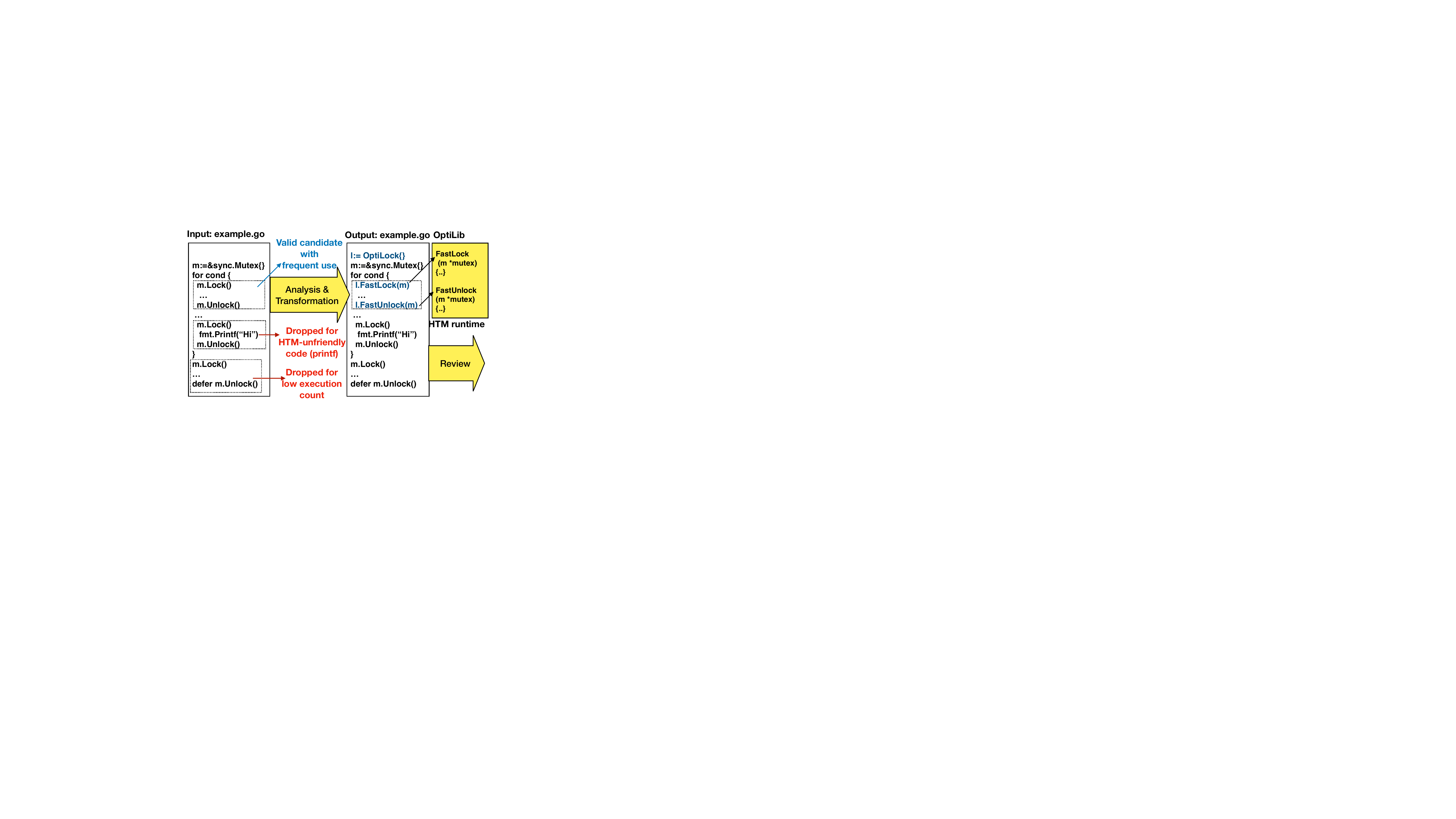}
    \caption{\small{\tool{} schematic diagram. Static analysis detects  three legal lock-unlock pairs in the input file \texttt{example.go}. The top one is a valid replacement candidate. The middle one is filtered since it contains I/O operations in its critical section. The bottom one is dropped due to the infrequent use via the information provided by profiles. The transformed code calls \lib{}, which executes the critical section via HTM. The resulting diff is given to the developer for review.}}
    \label{fig:schematic}
\end{figure}

Figure~\ref{fig:schematic} depicts our solution.
At a high level, our solution starts with using static analysis to identify candidate lock-protected critical sections to be instead protected by the HTM.
Then we filter out non-desirable candidates using both static analyses (e.g., to eliminate regions containing I/O operations) and dynamic analysis (to eliminate regions where the application of the HTM would not be beneficial based on profile data collected at runtime). Finally, we rewrite the code to have candidate regions use HTM constructs provided by the HTM library we developed instead of Go locks~\cite{gosyncPkg}.
\tool{} transformations are guaranteed to be safe; developer involvement is optional but highly recommended to let developers ultimately decide whether or not they want to use HTM. 

In summary, this paper makes the following contributions:
\begin{enumerate}[noitemsep,left=0pt,topsep=1pt]
\item We present the design and implementation of a framework for identifying lock-protected critical sections and select the best candidates for lock elision based on static analysis and execution profiles of Go programs.
\item We describe the source-to-source code transformation to replace mutual-exclusion locks in Go programs with HTM concurrency control constructs.
\item We introduce a library extending vendor-provided HTM primitives with intelligent features such as runtime contention management. 
Specifically, we devise a lightweight perceptron~\cite{jimenez2001dynamic,tarjan2005merging} that learns whether eliding a lock via HTM at a call site~\cite{graham1983execution} is beneficial at runtime.
\item We demonstrate the effectiveness of \tool{} for improving performance of real-life concurrent Go code by up to \speedup{}. 
\end{enumerate}

\section{Challenges}
\label{sec:challenges}

Locks are widely used in the real-world Go code and a significant amount of execution time can be spent waiting to acquire them~\cite{gobugs, goMutextCollapse, goLockContention,reduceLockCon2020, golocksStackOverflow, goMutexStarve, rwlockFootshoot} 
\footnote{A limited study we performed in a large-scale industrial setting using thousands of different Go services showed up to 30\% execution time being spent in lock-related code in certain Go programs; 5-10\% was quite common.}.
It is possible to replace a lock with a transaction that enables a critical section to be speculatively executed without actually holding the guarding lock.  With the support of the HTM, such replacements can result in significant speedups. However, there are several challenges in performing these replacements correctly and robustly, and ensuring that they deliver high performance reliably.

First, automatically and accurately matching a lock with its corresponding unlock operation to precisely identify critical sections is a complex problem.
Real-world programs can use locks with nesting intra- or inter-procedurally, which makes it significantly more involved. Additionally, certain lock-compatible instructions (e.g., IO and privileged instructions) will not work with HTM. A critical section including such instructions will not benefit from HTM.

Furthermore, Go provides a keyword that enables delaying lock release operation to all exit points of a function by prefixing the \texttt{Unlock()} operation with the \texttt{defer}~\cite{golangdefer} keyword\footnote{Any function can be deferred in Go.}.
It not only complicates matching an unlock with a lock operation, it may unnecessarily lengthen a critical section, which according to a synthetic benchmark we wrote shows performance degradation. 
A scan of 21 million lines of industrial Go code, which includes about 8000 \texttt{Unlock()} operations, shows that about 76\% are prefixed with the \texttt{defer} keyword. This indicates that handling \texttt{defer} statements is important.

Second, the Go language nuances~\cite{effectivego}  (e.g., pointer vs. value syntax, anonymous \mutex{} fields, lambda functions, etc.) make it non-trivial to transform lock-based code to HTM-based code.

Third, HTM has startup and commit overheads.
Even in non-concurrent code, where data-access conflicts do not happen, HTM can fail~\cite{singleCoreConflict},
 and locks may outperform HTM, particularly on tiny critical sections~\cite{RITSONonline}.

Fourth, the critical section size can be hard to estimate in general. 
If we make the conservative design choice and do not replace the lock if the critical section size is unknown, we can miss the opportunity to generate significant performance improvement. Thus, we need some runtime mechanism that can handle critical sections of arbitrary sizes with low overhead.

Fifth, when HTM aborts for a genuine data-access conflict, naively falling back to using a lock can be detrimental to performance~\cite{dice2009early, kleen2014tsx}. %
Deciding when and how to retry HTM-based executions or fall back to using fine-grained locks must be handled very carefully to avoid pathologies~\cite{dice2009early, kleen2014tsx, dice2008applications}.

Our tool, \tool{}, attempts to solve the above challenges.
\tool{} is an end-to-end system for improving the performance of lock-based Go code using HTM. 
We devise a sophisticated program analysis to identify lock-protected critical sections ($\S$~\ref{sec:analyzer}), support lock-to-HTM code transformation including non-trivial Go features ($\S$~\ref{sec:transform}), and develop an efficient HTM library to handle issues manifested at runtime ($\S$~\ref{sec:lib}).

\section{Related Work}

Herlihy and Moss proposed transactional Memory (TM)~\cite{herlihy1993transactional} in 1993 as an alternative to locks.
While locks proactively prevent two or more threads from concurrently accessing shared data, TM takes the opposite approach --- concurrent accesses are allowed as long as they do not conflict.
A lot of work has been done around both software and hardware implementations of transactional memory~\cite{tl2, ring-stm, mcrt-stm, htm-early, intel-tm, htm-blue, htm-blue-intel}, but only a few~\cite{stm-tires, tm-quake, ruan2014transactionalizing, karnagel2014improving} focused on evaluating the approach with real-life workloads, and none have done this for Go.

Intel's TSX extension of x86 instructions set~\cite{tsx} implementing HTM is of specific interest here as it underlies parts of our implementation. 
It is widely available in modern Intel CPUs and offers software interfaces providing subtly different functionality. The RTM (Restricted Transactional Memory) interface allows programmers to execute arbitrary code as a hardware transaction. All operations within a transaction have atomic execution behavior --- they all either appear to happen instantaneously or the entire transaction aborts and reverts the architectural state to before it was started. This can be trivially used to emulate the behavior of mutual-exclusion locks. In fact, this is precisely the kind of functionality that the HLE (Hardware Lock Elision) interface provides. However, HLE has been introduced mainly for backward compatibility with architectures that are not TSX-enabled and is not only very simplistic (e.g., with respect to contention management) but has also been shown to perform poorly compared to RTM~\cite{htm}. Consequently, our solution uses the RTM interface as the low-level implementation mechanism to build a comprehensive TM-based alternative for mutual-exclusion locks.

Lock elision, whether in software or hardware or a hybrid fashion, including gaining insights into them, has been extensively studied~\cite{Dalessandro2011, dice2008applications , Jacobi2012, htm-blue-intel, DiceTLE2016 , Afek2014, calciu2014improved, Dice2014, Dieguesusenix2014, Diegues2014, htm-blue-intel, Riegel2011, intel-tm, chapman2016hybrid, izraelevitz2016implicit, zheng2017exploiting, sousa2017fgscm, wang2019lightweight, KleenGNU, Pohlack2011}.
Our work uses many of those techniques; for example, the basic design of our runtime controller was inspired by Wang et al.~\cite{wang2019lightweight}. Additional possibilities to bring more solutions from the literature to the design and implementation of both our static analysis tool and runtime controller also exist.
Other attempts to use transactional memory for emulating mutual-exclusion locks exist as well~\cite{stm-single-lock-java, sle-java}, but they have to cope with higher overheads and semantics-related complications due to using the STM, they target the Java language whose synchronization lock-like primitives (i.e., \emph{monitors}) are easier to handle due to their lexical scoping and, most importantly, their evaluation is based exclusively on synthetic benchmarks.

\section{\tool{} Overview}
\label{sec:methodology}
A Go \mutex{} is a runtime object with  \texttt{Lock()} and \texttt{Unlock()} operations on it. Two (or more) critical sections guarded by the same \mutex{} will not execute concurrently.
When transforming locks into HTM, there are two possibilities.
\begin{enumerate}[noitemsep,left=0pt,topsep=1pt]
    \item A given \mutex{} guarding a set of critical sections is replaced with another object supporting operations analogous to \texttt{Lock()/Unlock()} but provided by the HTM. As a result, all critical sections previously guarded by the \mutex{} are now executed under HTM's control. 
    \item \texttt{Lock()/Unlock()} operations of the \mutex{} are replaced with their HTM equivalents on a per critical section basis. As a result, some critical sections for a given \mutex{} are still guarded by the same \mutex{}, while the others execute under HTM's control.
\end{enumerate}

The former is doable only if it is beneficial to transform all \texttt{Lock()/Unlock()} operations using a given \mutex{}, and the \mutex{} object is defined in the code that we are rewriting.
Assessing the benefit of transforming the \mutex{} object would require inspecting every critical section it protects. 
A ``may alias'' pointer analysis~\cite{Landi91,Hind2001} can answer such a question.
The ``all-or-none'' coarse-granularity of this approach makes it unattractive because the imprecision of pointer analysis  overapproximates the critical sections protected by a \mutex{}, disqualifying too many \mutex{}es from transformation.

This work adopts the latter approach, where we consider pairs of \texttt{Lock()/Unlock()} operations in the code for transformation, which provides fine-grained control over transformation.
This approach has to handle pairing a lock with its corresponding unlock and support interoperability of HTM (where the code is transformed) with locks (where the code is not transformed). This kind of interoperability is well-studied in the literature~\cite{locality-preserving-locks, Afek2015, htm-early, adaptive-hsle,kleen2014tsx} and is handled by our library.

Recall that input to \tool{} is the source code for a Go program/library along with its execution profiles  (as depicted in Figure~\ref{fig:schematic}). The output is a source code patch, where candidate \texttt{Lock()/Unlock()} operations are replaced with calls to a custom HTM library. 
\tool{} consists of the following key components:
\begin{itemize}[noitemsep,left=0pt,topsep=1pt]
    \item Analyzer: performs static analysis on the input program and collects lock-unlock pairs for transformation ($\S$~\ref{sec:analyzer}).
    \item Transformer: rewrites the program by replacing \texttt{Lock()/Unlock()} with \fastpair{}, which elide the lock using HTM ($\S$~\ref{sec:transform}).
    \item Adaptive runtime (\lib{}): implements HTM in Go and provides required runtime mechanisms including retry and rollback ($\S$~\ref{sec:lib}).
\end{itemize}

The source code patch choice, rather than a compiler transformation, is motivated by the desire to keep the developers in the loop.
Using HTM without developers' knowledge can prove unwelcome because developers often demand full visibility into their  programs.
Developers are becoming performance and variance sensitive~\cite{MaricqVariance2019,SuVarianceSC19,HuangVProf2017}, and an accidental regression can become hard to diagnose. %
As a side effect, the choice of source-code patch demands us to be surgical --- injecting large, complicated  HTM-handling boilerplate code is a non-starter. Consequently, we perform \texttt{Lock()/Unlock()} operations replacements with API calls to HTM logic hidden in the \lib{} open-source library and do so only in places where benefits of HTM are likely (e.g., we minimize the number of modified code locations using execution profiles).

\subsection{{\tool{} Guarantees and Limitations}}
\begin{itemize}[noitemsep,left=0pt,topsep=1pt]
\item \tool{} will transform properly synchronized code (i.e., where every lock operation will have a corresponding unlock operation) into the equivalent code without changing the code's behavior. 
Code not meeting this criterion will be either  not transformed, or transformed and its runtime behavior will be unchanged.
\item \tool{} considers only those lock-unlock pairs that seem to operate on the same lock within the same function --- inter-procedural \texttt{Lock()/Unlock()} operations are disregarded. 
Note, however, that in a critical section protected by \tool{} transformed lock can  make arbitrary function calls.
The requirement to have both \texttt{Lock()} and its matching \texttt{Unlock()} operation be present in the same procedure scope is only our implementation choice and pragmatic in nature. Over 70\% of the locks we inspected met this criterion. 
\item \tool{} makes no effort to identify critical sections or code reachability in the presence of reflection~\cite{goReflectPkg}.
\item \tool{}, as implemented, does not \emph{statically} detect HTM conflicts or capacity limitations (see $\S$~\ref{sec:analyzer} for the details).
\end{itemize}

\section{\tool{} Design and Implementation}
Before diving into the details of \tool's design and implementation, we define some common terminology.

\subsection{Terminology}
Go's \texttt{sync} package provides two kinds of shared memory objects: \texttt{Mutex} and \texttt{RWMutex}. 
\tool{} handles them both, but in the following sections, without the loss of generality, we will only use the term \mutex{} for simplicity.
From an HTM transformation viewpoint, an \texttt{RWMutex} is no different from a \texttt{Mutex}, except \texttt{RWMutex} offers additional APIs for read-only accesses.

A \emph{critical section} \cs{} is all code regions protected by a pair of lock and unlock operations on the same mutex object \texttt{m} --- the notation for calling lock/unlock operations on \texttt{m} is \texttt{m.Lock()}/\texttt{m.Unlock()} where \texttt{m} is referred to as a \emph{receiver}.
\emph{Lock-point}, abbreviated with letter $L$ (\emph{Unlock-point} abbreviated with letter $U$), is a static location in the code where the \texttt{Lock()} (\texttt{Unlock()}) function is invoked on a \mutex{}.
\textbf{\lup{}} is a set of $L$ and $U$ points.
\textbf{\lupair{}} is a candidate pair of one \emph{lock-point} paired with an  \emph{unlock-point}.
In the runtime context, fastpath/HTM-path means the use of HTM, and slowpath/fallback-path  means the use of the original lock.

We utilize the  Abstract Syntax Tree (AST), program Control Flow Graph~\cite{cooperbook2007} (CFG), and Static Single Assignment (SSA)~\cite{ssa-cfg} form of program representation prevalent in the compiler literature.
In a CFG, nodes are basic blocks~\cite{cooperbook2007} of straight-line code, and edges are control flow relationships among them. \tool{} first transforms the source code to the AST form (which is also used for code transformation as described in $\S$~\ref{sec:transform}) and then to the SSA form for CFG construction. %

\subsection{Analyzer}
\label{sec:analyzer}

The goal of the analyzer is to find as many \lupair{}s as possible. %
The \lupair{}s that protect HTM-incompatible critical sections (e.g., those including IO operations) must be pruned.
This filtering serves two purposes: it reduces the number of code changes and  non-beneficial HTM transformations.
Complicated lock usage patterns, several Go language quirks, and pointer imprecision complicate the static analysis.
A comprehensive call-graph analysis is vital because critical sections often contain function calls.

\textbf{Conflicts:} A sophisticated static analysis may detect whether transactions conflict. Answering this question, however, is unlikely to be valuable because developers typically do not use a lock if a conflict is impossible. 
Assuming conflicts happen, there is no easy way to statically determine whether transactions do not ``typically'' conflict. 
We do not try to solve this problem and leave conflict resolution to \lib{}. 

\textbf{Capacity:} 
Although one can perform static analysis to estimate the memory footprint of a critical section, it may not be possible if the bounds of a loop are unknown.
Also, without knowing the target architecture's HTM capacity, it would be premature to filter out candidate critical sections this way. We leave the capacity-related decisions also to \lib{}.

In the rest of this section, we, first, define the scope of our transformation ($\S$~\ref{sec:match}); then, describe the process of matching a lock with an unlock operation within a code region assuming no lock nesting and no function calls in a \cs{}; extend our analysis to include nested locks ($\S$~\ref{sec:nest});  expand the analysis scope to \cs{}s that may contain function calls ($\S$~\ref{sec:interproc}); detail special case of Go's \texttt{defer} statement ($\S$~\ref{sec:defer}); and finally discuss profile-based filtering ($\S$~\ref{sec:profile}).

\subsubsection{Scope of Transformation} 
\label{sec:scope}
To simplify the analysis, if a \texttt{Lock()/Unlock()} operation is executed in the middle of a basic block, we break such basic blocks in the CFG so that each \lp{} begins a new basic block and each \up{} ends a basic block.
A single-entry single-exit (SESE) region~\cite{JohnsonPLDI94} (simply \emph{region}) of a CFG is our smallest granularity of lock transformation.
A region is a subgraph of a CFG.
Control reaching any basic block in a region is guaranteed to have already executed a designated entry basic block; control leaving  from any basic block in the region is guaranteed to eventually pass through a designated exit basic block.

A function is the largest granularity of our lock transformation; a function always forms a region because all exits from a function are considered to go through a dummy basic block.
This choice is pragmatic in nature since \lupair{}s spanning multiple functions are uncommon. %

Regions can be nested within one another.
A Program Structure Tree (PST) organizes regions into a hierarchical tree~\cite{JohnsonPLDI94}.
We visit regions inside out from most-nested to least-nested. Appendix~\ref{sec:straightlinedom} describes the region identification and visiting strategies, which are not central to this paper.

\subsubsection{Matching \lupair{} in the Absence of Nested Locks}
\label{sec:match}
This subsection discusses analyzing a candidate region $R$.

\lup{} in $R$ may be operating on different locks, which should be pruned.
Some lock (unlock) operations may escape $R$, without a corresponding unlock (lock) operation in $R$, which should also be pruned. Below, we formalize these aspects.

\begin{figure}[t!]
\begin{minipage}{.45\linewidth}
\begin{lstlisting}[language=Go,label={lst:original2},caption=Original lock-based code.]

 m := &sync.Mutex{}
 m.Lock()
 m.Unlock()
\end{lstlisting}
\end{minipage}
\begin{minipage}{.45\linewidth}
\begin{lstlisting}[language=Go,label={lst:correct2},caption=Transformed HTM code.]
 l := OptiLock{}
 m := &sync.Mutex{}
 l.FastLock(m)
 l.FastUnock(m)
\end{lstlisting}
\end{minipage}
\end{figure}

\begin{Definition}[Points-to set ${\cal M}(L)$ of a Lock point $L$]
\label{def:ptsMutex}
Every \lp{} ($L$) operates on some receiver mutex pointer \texttt{p}.\footnote{At the source level \texttt{p} can be either a pointer or an actual object value, but at the SSA level it is always a pointer.} 
Such a mutex pointer may point to one or more mutex objects allocated in the program.
The set of all possible \mutex{} objects that \texttt{p} may point to in the program is the Points-to Set of $L$, denoted by ${\cal M}({L})$.
\end{Definition}

Similarly, the Points-to set of an Unlock point $U$ is  ${\cal M}(U)$.
We employ Anderson's flow-insensitive may-alias analysis~\cite{Andersen94programanalysis} to obtain  ${\cal M}({L})$ and  ${\cal M}({U})$ on the whole program.

\begin{Definition}[Downward Exposed Lock-point (\textbf{DELock})]
\label{def:lpup}
A \lp{}, $L$, with points-to set ${\cal M}(L)$, is downward exposed in region $R$, if there exists at least one path from $L$ to $R$'s exit without any \up{} on any mutex in ${\cal M}(L)$.
\end{Definition}

\begin{Definition}[Upward Exposed Unlock-point (\textbf{UEUnlock})]
\label{def:uplp}
An \up{}, $U$, with points-to set ${\cal M}(U)$, is upward exposed  in region $R$, if there exists at least one path from $R$'s entry to $U$ without a \lp{} on any mutex in  ${\cal M}(U)$.
\end{Definition}

\begin{figure}[!t]
\fbox{
\begin{minipage}{0.45\linewidth}
    \centering
    \includegraphics[width=.9\linewidth]{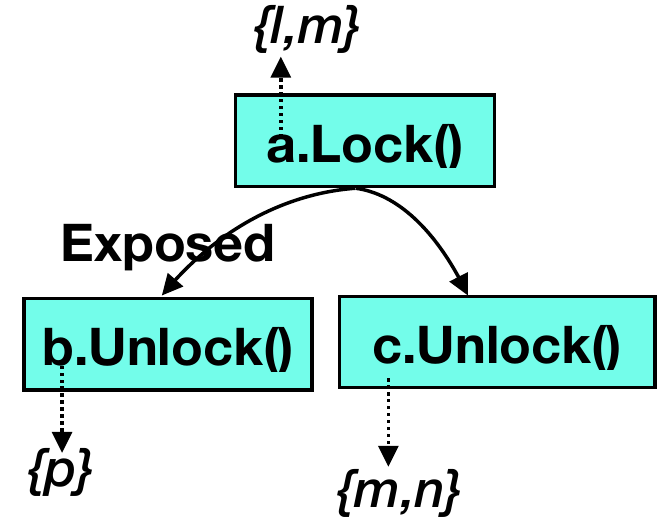}
    \caption{\texttt{a.Lock() is Downward Exposed.}}
    \label{fig:delock}
\end{minipage}}
\fbox{\begin{minipage}{0.45\linewidth}
    \centering
    \includegraphics[width=.9\linewidth]{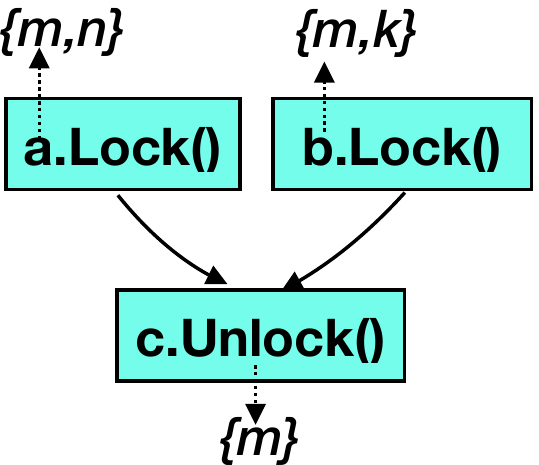}
    \caption{\texttt{c.Unlock()} is \textbf{not} Upward Exposed.}
    \label{fig:ueunlock}
\end{minipage}}
\end{figure}

\begin{figure}[!t]
\fbox{
\begin{minipage}{\linewidth}
    \centering
\includegraphics[width=.45\textwidth]{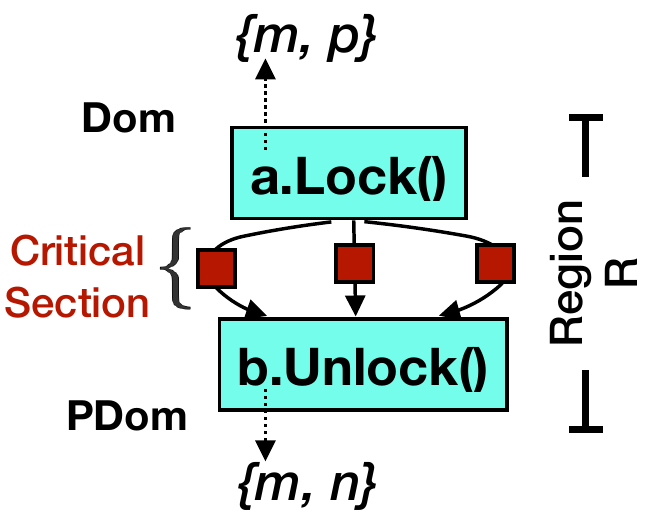}
    \caption{Region dominated by lock and post-dominated by unlock.}
    \label{fig:dompdom}
    \end{minipage}}
\end{figure}

$DELock$ identifies \lp{}s that \emph{definitely} do not have any corresponding \up{}s in some execution paths in $R$; and $UEUnlock$ identifies \up{}s that \emph{definitely} do not have corresponding \lp{}s in some execution paths in $R$.

Figure~\ref{fig:delock} exemplifies a downward exposed \lp{}.
Mutex pointer $a$'s points-to set $\{l,m\}$, has an empty intersection with $b$'s points to set $\{p\}$; although it has a non-empty intersection with $c$'s points-to set $\{m, n\}$.
Figure~\ref{fig:ueunlock} exemplifies an \up{} that is \emph{not} upward exposed.
Mutex pointer $c$'s points-to set $\{m\}$, has non-empty intersection with $a$'s  points-to set $\{m, n\}$ and $b$'s points-to set $\{m,k\}$.

We eliminate all $DELock(R)$ and $UEUnlock(R)$ from the transformation in $R$. 
The remaining \lp{}s in $R$ are the complement of $DELock(R)$, which is denoted by $\overline{DELock(R)}$. 
Similarly, the remaining \up{}s in $R$ are the complement of $UEUnlock(R)$, which is denoted by $\overline{UEUnlock(R)}$.

\begin{Definition}[Feasible-HTM-Pair]
\label{def:feasible}
Let ${L} \in \overline{DELock(R)}$. %
Let ${U} \in \overline{UEUnock(R)}$. %
$L$ and $U$ form a feasible HTM  pair if all of the following conditions are true,\\
(1) ${\cal M}({L}) \cap {\cal M}({U}) \ne \phi$, \\
(2) $\big(L$ \textsc{Dom} $U\big) \bigwedge  \big(U$ \textsc{PDom} $L\big)$,\\
(3) The critical section $C \subseteq R$ guarded by $L$ and  $U$ contains no LU-point $X$ such that ${\cal M}({X})  \cap \big({\cal M}({L}) \cup {\cal M}({U})\big) \ne \phi$, and \\
(4) $C$ contains no HTM-unfriendly instructions.

\end{Definition}

Condition (1) filters out those \lup{} that are guaranteed to be operating on different \mutex{}es.

Condition (2) filters out infeasible control flows where unlock happens before lock and vice-versa.
\textsc{Dom} and \textsc{PDom}~ respectively represent dominator~\cite{cooperbook2007} and post-dominator~\cite{cooperbook2007} relationships in a CFG.
Figure~\ref{fig:dompdom} shows an example, where
all paths from \lp{} \texttt{a.Lock()} are post-dominated by \up{} \texttt{b.Unlock()}, whose all incoming paths are dominated by \texttt{a.Lock()}.
Additionally, the set-intersection of the points-to set of mutex pointers $a=\{m, p\}$ and $b=\{m, n\}$ is non-empty.
Any Feasible-HTM-Pair on $L$ and $U$, forms an SESE-region by itself, where the entry basic block has $L$ as its first instruction and the exit basic block has $U$ as its last instruction.
Condition (2) intuitively finds correct candidate \lupair{}s in the absence of nested locks because if a  lock operation $L$ is performed on every path reaching any code in $C$ and an unlock operation $U$ is performed on every path exiting $C$, then $LU$ must be operating on the same \mutex{}.
 We further discuss our choice of \textsc{Dom/PDom} relationships in Appendix~\ref{sec:dompdom}.

Condition (3) ensures that if we match an $L$ with a $U$, there does not exist another \lp{} or \up{} in the same region that \emph{may} operate on a \mutex{} in the same points-to set as that of $L$ or $U$. The next subsection elaborates on lock nesting.

Condition (4) is an obvious requirement to ensure HTM does not abort.
A region is unsafe if it contains any IO instructions. %

Since we use ``may alias'' to match a \lp{} with \up{}, it is possible (but less likely) for our transformation to pair a lock with an unlock that may be operating on two different mutex objects at runtime. 
However, at runtime, we can obtain and memorize the address of the mutex object used at the \lp{}, and compare it against the mutex object offered to the runtime at the \up{}.
In case of an address mismatch of the mutex objects used in the same \lupair{}, we can abort the transaction and revert to a safe state and fall back to using the locks. 
A mismatch is impossible without nested locks because of the dominance and post-dominance relationship between the lock and unlock in an \lupair{}.

\subsubsection{Lock Nesting}
\label{sec:nest}

Go supports nested locks, but reentrant~\cite{reentrantWiki} locks are not allowed.
Condition (3) in Definition~\ref{def:feasible} allows nested locks but demands that they operate on disjoint \mutex{} objects. %

HTM via Intel TSX allows nesting: if a nested transaction succeeds, hardware does not commit it until the outermost transaction commits. 
If a nested transaction fails, the control jumps to the starting code address of the nested transaction. 
This facility allows us to safely transform locks into HTM even when they are nested.

Condition (3) in Definition~\ref{def:feasible} disqualifies a candidate \lupair{} from the transformation in region $R$ if there exists any other lock or unlock point whose lock/unlock operation \emph{may be} operating on the same mutex as those in the \lupair{}.

As an example, in Listing~\ref{lst:locknest}, assume the mutex pointers $a$ and $b$ point to the same points-to set. When inspecting the ``inner region'', we find only one \lupair{}, which obeys all Feasible-HTM-Pair conditions in Definition~\ref{def:feasible}. Consequently, the lock usage on $b$ in the inner region can be transformed to HTM.
When inspecting the ``outer region'', however, we see conflicting \lup{}, and hence the locking operations on $a$ will not be transformed. The resulting transformed code is shown in Listing~\ref{lst:locknestHTM}, which is correct.

\begin{figure}[t!]
\begin{minipage}{.60\linewidth}
\begin{lstlisting}[language=Go,label={lst:locknest},caption=Nested Locks.]
 a.Lock() //outer region start


 b.Lock()  // inner region start
 b.Unlock() // inner region end
 
 a.Unlock() //outer region end
\end{lstlisting}
\end{minipage}
\begin{minipage}{.35\linewidth}
\begin{lstlisting}[language=Go,label={lst:locknestHTM},caption=HTMized.]
 a.Lock()
 
 l := OptiLock{}
 l.FastLock(b)  
 l.FastUnlock(b)
 
 a.Unlock()
\end{lstlisting}
\end{minipage}
\end{figure}

This approach complicates hand-over-hand locking~\cite{ChakrabartiAtlas2014, Kelly2020}, sometimes used in the concurrent linked-list traversal, shown in Listing~\ref{lst:hohlock}.
As before, assume all four \lup{} have a non-empty intersection of their points-to sets.
When inspecting the inner region, the \lupair{} \texttt{b.Lock()} and \texttt{a.Unlock()} passes all tests in Definition~\ref{def:feasible}. 
Hence, they will be, incorrectly, paired and transformed to use HTM, as shown in Listing~\ref{lst:hohlockHTM}. 
This transformation violates the programmer's intention.
Subsequently, when visiting the outer region, condition (3) is violated, and hence the outer \lupair{} will not be transformed.
One could have discarded the transformation of the inner region when the conflict is visible in the enclosing region. However, we cannot distinguish this incorrect pairing from the correct pairing in the previous case.
Our solution is to always apply the transformations on the candidates found in inner regions, and handle mismatches at runtime via HTM aborts iff executing on the fastpath.
As mentioned at the end of $\S$~\ref{sec:match}, a mismatch is easy to recognize at runtime by, first, making \texttt{FastLock()} store the address of the \mutex{} used at the \lp{} in a field in \optiLock{} and, second, checking whether the \mutex{} passed to \texttt{FastUnlock()} is the one present in \optiLock{}.
The transactional abort is needed (and possible) only on the fastpath.
 Appendix~\ref{sec:locknesting} details the correctness of transforming nested locks into HTM via \tool{}.

\begin{figure}[t!]
\begin{minipage}{.60\linewidth}
\begin{lstlisting}[language=Go,label={lst:hohlock},caption=Hand-over-hand lock.]
 a.Lock() //outer region start


 b.Lock()  // inner region start
 a.Unlock() // inner region end
 
 b.Unlock() //outer region end
\end{lstlisting}
\end{minipage}
\begin{minipage}{.35\linewidth}
\begin{lstlisting}[language=Go,label={lst:hohlockHTM},caption=HTMized.]
 a.Lock()

 l := OptiLock{}
 l.FastLock(b)  
 l.FastUnlock(a)

 b.Unlock()
\end{lstlisting}
\end{minipage}
\end{figure}

\subsubsection{Critical Sections with Function Calls} 
\label{sec:interproc}

When the critical section protected by a candidate \lupair{} contains function calls, we need to extend the analysis beyond the current function.
Conditions (1) and (2) in Definition~\ref{def:feasible} are local to $R$.
Conditions (3) and (4) require inter-procedural analysis.

We need to ensure that the transitive-closure of all code regions protected by a candidate  \lupair{}, including the blocks reachable via function calls,  neither contains any HTM-unfriendly instructions nor contains any \lup{} whose points-to set may overlap with the points-to sets of $L$ or $U$. 
Otherwise, we discard the candidate \lupair{}.

To accomplish this, we first build a static call graph using rapid type analysis~\cite{RTABacon, gorta}.
Next, we precompute summary information for each function on its own without its transitive closure of function calls; the summary contains 
(a) the fit of the function for HTM based-execution (i.e., no HTM-unfriendly instructions),  and
(b) the union of all points-to sets of all \lup{} in the function, denoted by $\cal {P}$.

For a candidate \lupair{} meeting all the conditions in Definition~\ref{def:feasible}
within the region $R$, we proceed to do inter-procedural analysis.
Let  $F^*$  be the transitive closure of all the function calls invoked inside the critical section %
$C \subseteq R$ protected by a candidate \lupair{}.
\lupair{} is discarded if  (a) $\exists F \in F^* s.t.\ F$'s summary contains  HTM-unfriendly instructions or (b)  $\exists F \in F^* s.t.$  ${\cal P} \cup ({\cal M}(L) \cup {\cal M}(U)\big) \ne \phi$.
The former is simply the condition (4) expanded to all functions, and the latter is condition (4) expanded to all functions.
We note that nested locks discussed in $\S$~\ref{sec:nest} can be in different functions.

\subsubsection{The \texttt{defer} Statement} \label{sec:defer}
The \texttt{defer}~\cite{golangdefer} statement in Go, introduced in $\S$~\ref{sec:challenges}, needs special attention.
Go defers the execution of functions prefixed with the \texttt{defer} keyword to the calling function's return point.
The presence of \texttt{defer Unlock()} complicates our CFG-based dominance/post-dominance analysis.
Deferred unlocks extend the critical sections till function exit points.
Listing~\ref{lst:defer} shows a legal Go code, where the \texttt{defer m.Unlock()} appears even before the call to \texttt{m.Lock()}. 
Condition (2) in Feasible-HTM-Pair will treat this pair as an invalid candidate for transformation because the \lp{} does not dominate the \up{}.

We address this case by interpreting \texttt{defer m.Unlock()} in a CFG by (a) introducing a synthetic \texttt{m.Unlock()} statement at the end of each basic block that returns control from the function, and (b) discarding the presence of \texttt{m.Unlock()} in its original position during the analysis.
This allows us to reuse the previously described dominance relationship.
During transformation, however, \tool{} simply replaces a \texttt{defer Unlock()} with a \texttt{defer FastUnlock()} in its original place, as shown in Listing~\ref{lst:deferHtm}.

Multiple \texttt{defer} calls are executed in a last-in first-out (LIFO) order of encountering the \texttt{defer} statement at runtime. This complicates the dominance and post dominance relationship; for simplicity, we discard any function that contains multiple \texttt{defer Unlock()} statements. 
We found none in the packages used in our evaluation.

\subsubsection{Profiles to Filter Hot Critical Sections}
\label{sec:profile}
Profiling is a built-in feature in Go, which takes callstack samples via timer~\cite{pprof} or hardware performance counter~\cite{chabbiPprof} interrupts.
One can take CPU profiles of a go program either at launch time by simply passing a \texttt{-cpuprofile} flag or from an already running program, for a specified duration, by accessing an exposed profiling port.

Go profiles are in the \texttt{pprof} format.

The pprof Go package~\cite{pprof} allows us to programmatically navigate the callstack samples presented as weighted call graphs, where the nodes represent functions and edges represent caller-callee relationships. The functions are annotated with their inclusive and exclusive execution times.

When profiles are available, we use them to filter the regions where negligible execution time is spent, even before applying the static analysis.
In fact, this is the first filtering step we perform before making the aforementioned \lupair{} identification.
We treat any critical section (including the entry and exit) where the aggregated execution time is less than 1\% of the total execution time as insignificant.

\begin{figure}[t!]
\begin{minipage}{.45\linewidth}
\begin{lstlisting}[language=Go,label={lst:defer},caption=defer Unlock.]
func DeferExample() {

  m := &sync.Mutex{}
  defer m.Unlock() 
  m.Lock()
  // critical section
}
\end{lstlisting}
\end{minipage}
\begin{minipage}{.45\linewidth}
\begin{lstlisting}[language=Go,label={lst:deferHtm},caption=HTMized.]
func DeferExample() {
  l := OptiLock{}
  m := &sync.Mutex{}
  defer l.FastUnlock(m) 
  l.FastLock(m)
  // inside HTM
}
\end{lstlisting}
\end{minipage}
\end{figure}

\subsection{Transformer}
\label{sec:transform}

\begin{figure}[t!]
    \centering
    \includegraphics[width=.7\linewidth]{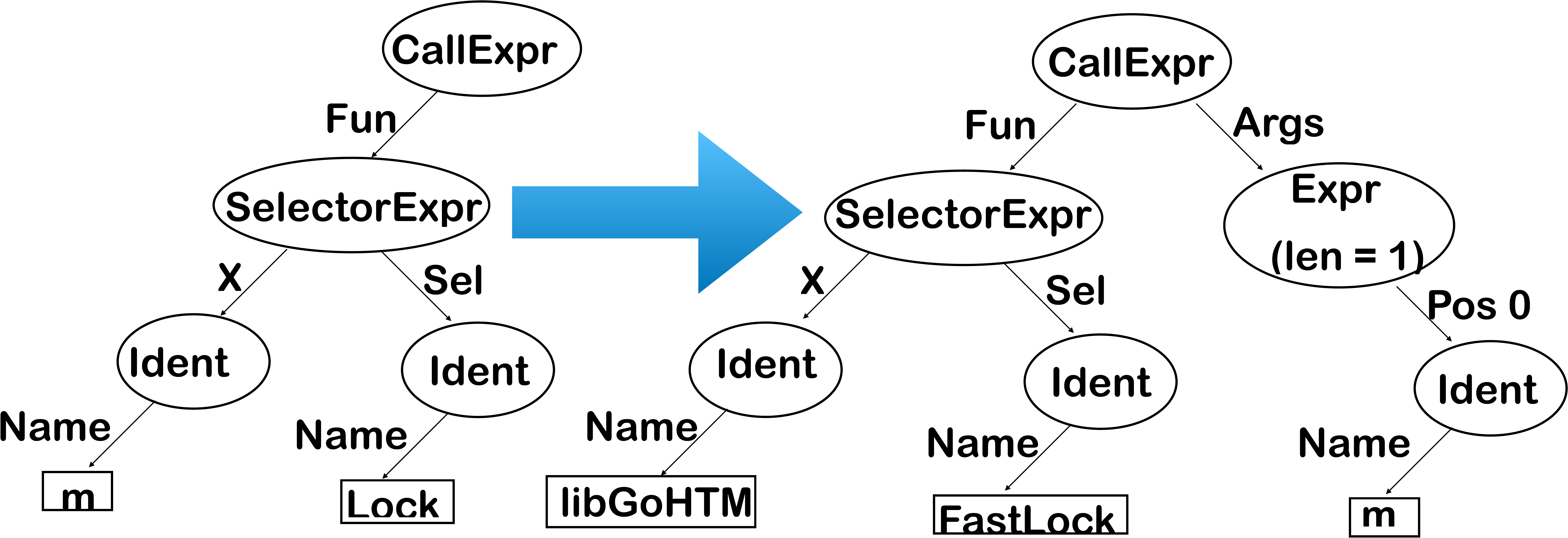}
    \caption{Example of AST transformation from \texttt{m.Lock()} to \texttt{\lib{}.FastLock(m)}. Some AST nodes are omitted for brevity.}
    \label{fig:transform}
\end{figure}

Since our end product is a code patch, we perform the transformation on the AST form of the program.
Go AST can be serialized into source code via Go \texttt{format}~\cite{goFormatPkg} package.
To this end, the transformer maps the candidate set of \lupair{} operations found during the SSA-based analysis phase  (described in $\S$~\ref{sec:analyzer}) to AST nodes~\cite{goASTUtilPkg}.
It then replaces the \lupair{} operations with calls to \texttt{FastLock()/FastUnlock()} in \lib{}. 
It also passes the original \mutex{} object as a pointer to the calls to \texttt{FastLock()/FastUnlock()} since the underlying lock object is necessary for lock elision (fastpath) as well as for slowpath.
Figure~\ref{fig:transform} shows an example AST transformation.
The transformation itself is mechanical but challenging. In the following sections, we discuss several Go features that pose special challenges in transforming the AST.

\paragraph{Go pointer vs. value:}
Go syntax does not distinguish accessing a field of a composite type (e.g., \texttt{struct}) via an object-pointer or an object-value; both use the same  AST \texttt{dot} operator as exemplified in Listing~\ref{lst:ptrVSval}.
However, \texttt{FastLock()} and \texttt{FastUnlock()} must receive a pointer to the \mutex{} object to perform the elision correctly.
Hence, if the \lupair{} uses a \mutex{} value, its address needs to be passed to \texttt{FastLock()/FastUnlock()}, and if the \lupair{} uses a \mutex{} pointer, it should be passed as is.

\begin{figure}[t!]
\begin{minipage}{.45\linewidth}
\begin{lstlisting}[language=Go,label={lst:ptrVSval},caption=Both Mutex pointer \texttt{m} and Mutex value \texttt{n} invoke Lock/Unlock using the same \texttt{dot} dereferencing operator.]

// pointer form
m := &sync.Mutex{} 
m.Lock()
m.Unlock()

// value form
n := sync.Mutex{}
n.Lock()
n.Unlock()
\end{lstlisting}
\end{minipage}
\begin{minipage}{.45\linewidth}
\begin{lstlisting}[language=Go,label={lst:ptrVsValHTM},caption=GOCC transformation passes \texttt{m} as-is to FastLock()/FastUnlock() but \texttt{{\&}n} to FastLock()/FastUnlock().]
l := Optilock{}
// pointer form
m := &sync.Mutex{} 
l.FastLock(m) 
l.FastUnlock(m)

// value form
n := sync.Mutex{} 
l.FastLock(&n)
l.FastUnlock(&n)
\end{lstlisting}
\end{minipage}
\end{figure}

We address this issue in the transformer by querying the type information~\cite{golangTypes} for each receiver object 
subject to transformation.
If the receiver identifier is a \texttt{Mutex} value type, we insert the additional \texttt{address-of} operator before the \mutex{} identifier in the AST and pass it to \texttt{FastLock()/FastUnlock()}.
If the receiver identifier is a pointer to a \texttt{Mutex} type, we pass it as is.

\paragraph{Go anonymous fields:} Go allows programmers to define a struct that has fields without names.
For instance, Listing~\ref{lst:annoyField} shows a struct \texttt{AStruct} that has an anonymous \texttt{*sync.Mutex} field. 
Operations on this anonymous mutex are performed by simply using the name of the enclosing struct variable as shown on Line~\ref{lst:annoyFieldref}.
Hence, our transformation needs to be cognizant about whether the  original \lupair{} operations are performed on an anonymous mutex.

By inspecting the type information~\cite{golangTypes} of the \emph{access path}~\cite{Johannes2015} used to invoke the lock/unlock operation in the AST, we determine whether or not the operation is performed on an anonymous mutex field.
Upon determining the operations to be on an anonymous mutex, we pass the address of the anonymous \texttt{Mutex} to \lib{} by simply suffixing the operation access path with \texttt{Mutex} as shown in Listing~\ref{lst:annoyFieldGOCC}, Line~\ref{lst:annoyFieldGOCCref} (where access path simply consists of variable \texttt{a}). 
This transformation composes with the previously described pointer vs. value operations.

\begin{figure}[t!]
\begin{minipage}{.45\linewidth}
\begin{lstlisting}[language=Go,label={lst:annoyField},caption=Locking on an unnamed field of a struct.]
type AStruct struct {
  x int //not anonymous
  *sync.Mutex //anonymous
}
func main{} {

  a := AStruct{}
@\label{lst:annoyFieldref}@  a.Lock()
  a.Unlock()
}
\end{lstlisting}
\end{minipage}
\begin{minipage}{.45\linewidth}
\begin{lstlisting}[language=Go,label={lst:annoyFieldGOCC},caption=Unnamed mutex transformed to HTM.]
type AStruct struct {
  x int // not anonymous
  *sync.Mutex //anonymous
}
func main{} {
  l := OptiLock{}
  a := AStruct{}
@\label{lst:annoyFieldGOCCref}@  l.FastLock(a.Mutex)
  l.FastUnlock(a.Mutex)
}
\end{lstlisting}
\end{minipage}
\end{figure}

\paragraph{Anonymous goroutines:} Go supports anonymous goroutines~\cite{goClosure}, which are nested inside other functions as shown in Listing~\ref{lst:lambdaFunction}; these goroutines run concurrently.
Our transformation introduces a new \optiLock{} variable in transformed functions.
\optiLock{}'s declaration should be in the scope that encloses both Lock and Unlock operations, but it should not be shared by other concurrent executions because it maintains goroutine-specific state.
Hence, we make \optiLock{} a variable in the stack of each goroutine.
We add the declaration to the innermost function body as shown on line~\ref{lst:lambdaFunctionGOCCOpt} in Listing~\ref{lst:lambdaFunctionGOCC}.
A  bottom-up AST walk from \lupair{} to be transformed allows us to easily detect the innermost enclosing anonymous function scope. 

\begin{figure}[t!]
\begin{minipage}{.45\linewidth}
\begin{lstlisting}[language=Go,label={lst:lambdaFunction},caption=Anonymous goroutines create concurrent units of execution on anonymous functions.]
 m := &sync.Mutex{}
 for i:=0;i<10;i++ {
  go func() {

     m.Lock()
     // CS
     m.Unlock()
   }()
  }
//wait all
\end{lstlisting}
\end{minipage}
\begin{minipage}{.45\linewidth}
\begin{lstlisting}[language=Go,label={lst:lambdaFunctionGOCC},caption=The \optiLock{} needed for the transformation should be  placed in the innermost function scope.]
 m := &sync.Mutex{}
 for i:=0;i<10;i++ {
 go func() {
@\label{lst:lambdaFunctionGOCCOpt}@     l := OptiLock{}
     l.FastLock(m)
     // CS
     l.FastUnlock(m))
   }()
  }
//wait all
\end{lstlisting}
\end{minipage}
\end{figure}

\subsection{Adaptive HTM Runtime: \lib{}}
\label{sec:lib}

\lib{} implements all the intelligent runtime control needed to perform HTM in lieu of locks.
It is in charge of starting and committing transactions in critical sections, as well as falling back to the lock when necessary.
It is responsible for inter-operating with locking operations on the same mutex that may not be transformed to use HTM.
In the event of aborts, it is responsible for determining the cause of the abort and deciding whether and how many times to retry.
If, accidentally, the code rewriting matches \lp{} with a programmer-unintended \up{}, \lib{} is responsible for detecting and recovering from the mistake.
Finally, it is responsible for understanding and dynamically adjusting to changing contention.

We implement \lib{} using TSX~\cite{tsx} for Intel platforms. 
\lib{} is carefully implemented to ensure correctness under all circumstances. Equally important, it is implemented with the utmost attention to performance. Every instruction and its placement are carefully planned to minimize any overhead of its own.
\lib{} uses Intel RTM; it does not use the Hardware Lock Elision (HLE)~\cite{htm} because it does not provide the fine-grained control we need.

\lib{} introduces a data structure: \optiLock{}, which has two fields: a boolean named \texttt{slowPath} and a \texttt{*sync.Mutex} named \texttt{lkMutex}.
\texttt{slowPath} is set to true if the lock is not elided at runtime.
 \texttt{lkMutex} always holds the address of the fine-grained lock being elided.
\optiLock{} supports \texttt{FastLock()/FastUnlock()} operations, both need a \texttt{*sync.Mutex} argument, which is the mutex receiver object being elided at the original call sites of \texttt{Lock()/Unlock()}, respectively.

The \texttt{FastLock()} implementation uses sophisticated mechanisms described previously by others~\cite{locality-preserving-locks, kleen2014tsx, Afek2015} to interoperate slow and fast paths concurrently.
Stated succinctly, the \texttt{FastLock()}, waits for the fine-grained lock to be available before starting the hardware transaction; on starting a transaction, it first checks if the fine-grained lock is already held and unconditionally aborts if it is already held; if it is not held, the act of checking adds the lock word to the transaction read-set, and hence, if a concurrent execution on the slowpath  acquires the same lock during the transaction, the fastpath immediately aborts ensuring mutual exclusion. 
Any two threads in the fastpath can run concurrently as long as they do not conflict in their memory accesses.

Reading the internals of the original \texttt{sync.Mutex} object is straightforward and has no performance penalty; \texttt{FastLock()} simply de-references the first word of the \mutex{} pointer passed into the function, which contains the lock status.

The \texttt{FastUnlock()}, based on \texttt{slowPath} value, either commits the transaction or invokes the unlock on the mutex object.
It also safeguards against accidental incorrect code patches by ensuring that the mutex object passed into \texttt{FastLock()} and stored in the  \texttt{lkMutex} field of \optiLock{} matches the mutex object presented to \texttt{FastUnlock()}. In case of a mismatch, \texttt{FastUnlock()} restores safety by aborting the transaction (iff on fastpath), and subsequently enforces the slowpath.

\subsubsection{Dynamic adjustment via perceptron}
It would not be fruitful to attempt HTM if doing so has already proved to be unsuccessful for whatever reason.  
\tool{} learns and adapts to HTM behavior and decides whether to use HTM for the already transformed \lup{}, the fallback being the original lock.
For this purpose, \tool{} uses a featherlight, hardware-inspired “hashed perceptron”~\cite{tarjan2005merging}.

The hashed perceptron predictor hashes feature weights into one or more tables. 
Then at the prediction time, it uses indexes to access feature weights from the tables and adds up all the relevant weights.
If the sum exceeds a threshold, the prediction will be regarded as positive (e.g., "HTM should be taken").
Otherwise, the result will be viewed as negative (e.g., "HTM should not be used").
The weights will be updated based on the correctness of the predictions.

If operations on a given \mutex{} have been HTM-friendly/unfriendly, we want to utilize this information.
Similarly, if a code location has been HTM-friendly/unfriendly irrespective of the \mutex{} used, we want to use this information as well. 
Hence, the two input features for the perceptron are the \mutex{} and the calling context~\cite{Ammon97,graham1983execution} of lock/unlock invocation.
The address of the \mutex{} serves as the \mutex{} feature, and the address of \optiLock{} serves as a unique identifier for the calling context feature.
Updating the same perceptron weight for the \mutex{} feature by different goroutines would create a conflict (and potentially a performance collapse). Hence, we instead XOR the \mutex{} address with the address of the \optiLock{} to produce a conflict-free feature input.

Our perceptron implementation creates two 4K-entry arrays as the global weight tables (\gwt{}).
The weights take an integer number ranging from -16 to 15.
At runtime,  \texttt{FastLock()} and \texttt{FastUnlock()} functions index into \gwt{} by taking the lower-12 bits of the two features.
Perceptron operations are done outside the transaction.
The updates and reads from \gwt{} are lock-free but racy --- perfection is not required here, but high-performance is necessary.
Experiment results from $\S$~\ref{eval:perceptron} show the effectiveness of perceptron learning in protecting against poor HTM performance.

\begin{table*}[t!]
    \centering
    \begin{tabular}{c}
    \includegraphics[width=.9\linewidth]{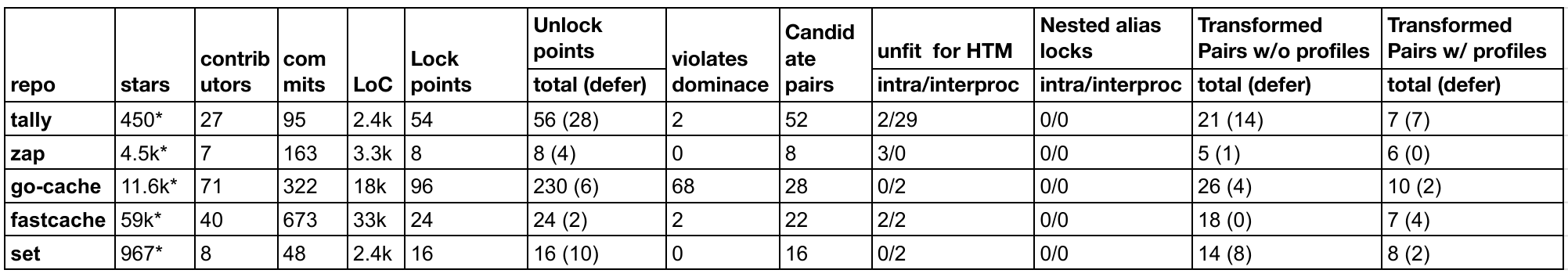}
    \end{tabular}
    \caption{Go package characteristics and their behavior using \tool{}}
\label{tab:benchs}
\end{table*}

\textbf{Perceptron weight update:} 
Perceptron weight updates happen in the \texttt{FastUnlock()} function after successfully finishing the critical section, whether on fast and slow path.

If the perceptron decides that the lock should be used, there will be no update to the weights as the lock will always succeed.
When the perceptron indicates to use the HTM and the execution finishes on the fastpath, the corresponding weight in the cell will be increased
(because the perceptron makes a correct decision, it should be encouraged to use the HTM more frequently).
On the other hand, if perceptron determines to use HTM, but HTM fails and falls back to slowpath, the weights will be decreased
(because HTM does not work for the current call, perceptrons should be penalized for incorrect recommendation to improve future predictions).

\textbf{Weight decay:} We keep a counter, in each cell in \gwt{}, to record the number of lock calls that go to the slowpath directly as a result of perceptron decision. 
If a lock has been used consecutively for a certain number of times and exceeds the threshold, we reset the weight of the perceptron cell and subsequently try HTM. 
Without this reset, perceptron would get stuck on the slowpath, preventing the benefits of the HTM execution in the future.
We set this threshold to 1000 continuous decisions. %
Appendix~\ref{sec:optilocalgo} summarizes our \texttt{FastLock()/FastUnlock()} implementations including the perceptron logic.

\subsubsection{Alleviating HTM overhead} 
HTM brings overhead for very short critical sections as described in $\S$~\ref{sec:challenges} above, even under single-core execution.
\lib{} avoids using HTM if it recognizes a single OS-thread in a Go process.
\lib{} employs \texttt{runtime.GOMAXPROCS(0)} API for this purpose.

\section{\tool{} Evaluation}
\label{sec:eval}
We evaluate \tool{} on an 8-core ($\times$2-way SMT~\cite{SMT95}) Intel Coffee Lake CPU with a total 32GB memory, running Linux \texttt{5.4.0}. 
The CPU has 32KB L1I and L1D cache, 256KB L2 cache, and 16MB L3 cache.
The Go version is 1.15.2. 

Table~\ref{tab:benchs} shows the list of applications and libraries we employ. 
In the absence of standard benchmarking for Go, we selected packages that are popular open-source Go projects (column 2 in Table~\ref{tab:benchs}), focus on high performance, utilize lock-based Go concurrency, and provide thread-safe APIs.
In particular, Zap and Tally are foundational logging and metrics collection packages used in production go programs by many organizations.
Additionally, since we evaluate the projects using their own benchmark suites (more on this below), we only selected projects that feature concurrent benchmarks or whose benchmarks could be straightforwardly converted to be concurrent.

From a static analysis viewpoint, we see that all applications contain several locks. 
Defer unlocks are common (column 7). 
The ``violates dominance'' column shows how many \lup{} were discarded since they did not meet the dominance relationship.
The number is typically low except for \texttt{go-cache}, which has several functions with the repeating pattern of  unlocks that do not post-dominate the candidate lock.
The ``candidate pairs'' column shows how many \lup{}  remain for further analysis.
Each column to the right progressively shows the reasons for which a candidate \lupair{} was rejected. 
Rejection due to nested aliased locks is not found in these packages.
The second-to-last column shows how many \lupair{}s were finally rewritten to use \optiLock{}, including how many of them contain \texttt{defer Unlock()}.
The last column shows the numbers after we retain only those locks where the functions contain at least 1\% of execution time in execution profiles.
Overall, \tool{} transforms several \lupair{}s in each application.
Using profiles significantly reduces the number of transformed \lupair{}s.

We run all the benchmarks within each repository five times and report the median.
We believe the benchmarks accompanying the code best represent its desired characteristics.
As some benchmarks are written for a single thread setting, we rewrite them to introduce concurrency to utilize HTM-enabled parallelism fully. 
We adopt the standard testing package from Go\cite{golangTesting}, which runs each benchmark for a certain amount of time and reports the throughput as nanoseconds per operation.
We wrap the benchmark codes with RunParallel~\cite{golangRunParallel} helper function to get parallel performance if it was not already done so. 
Using more CPU cores, ideally, increases throughput (i.e., reduces average nanoseconds per operation).
Then  we compare the throughput from locks vs. HTM --- a positive percentage means \tool{}'s rewrite did better, and a negative percentage means the baseline did better.
We vary the number of CPU cores available for benchmarks from 1 to 8. 
Unlike HPC codes that run on all cores on a server, Go services often use 2-4 cores.
\begin{figure*}[!t]
\begin{minipage}{.5\linewidth}
    \centering
    \includegraphics[width=.9\linewidth]{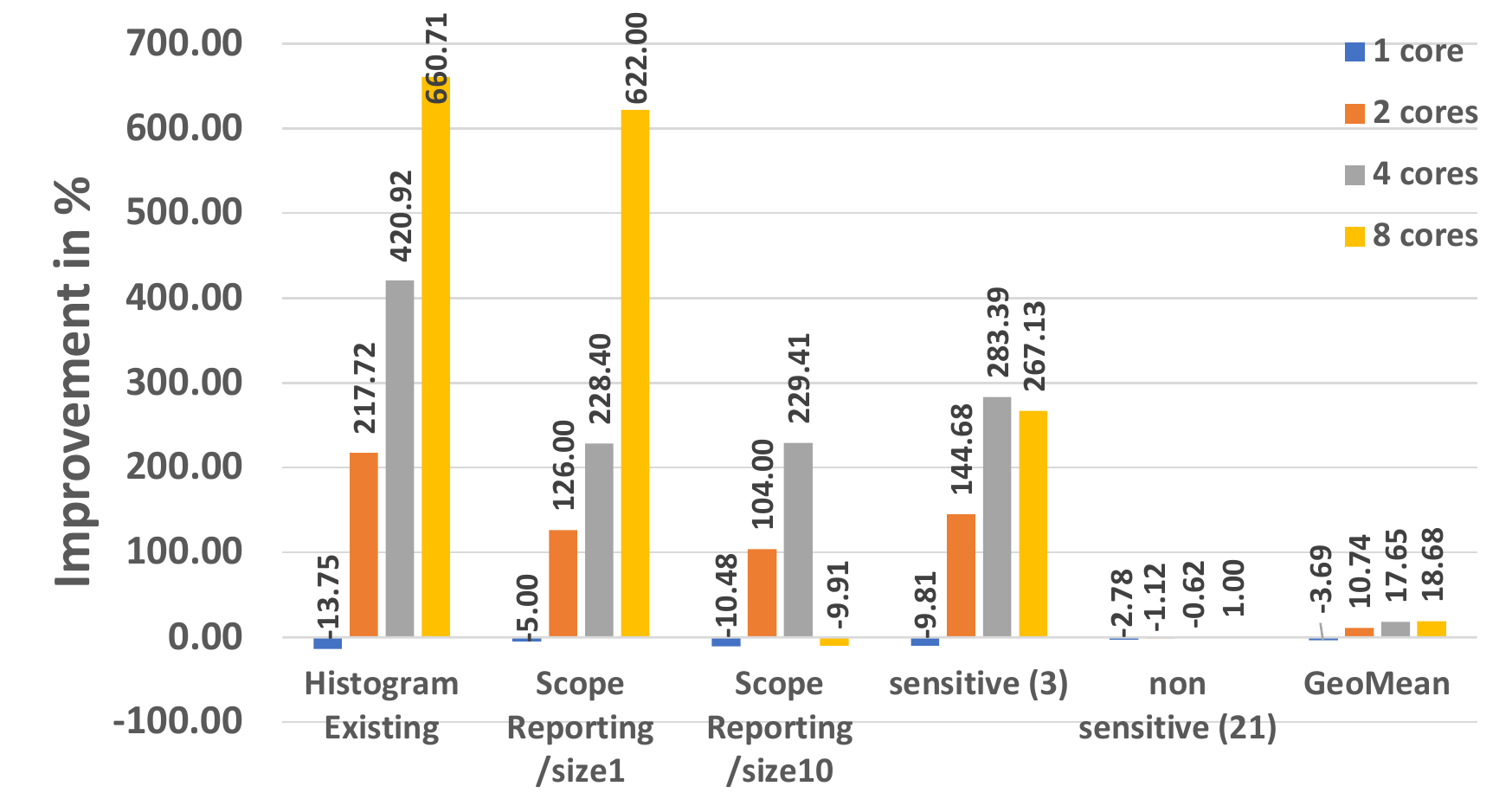}
    \caption{Results on Tally with different core numbers.}
    \label{fig:tally}
\end{minipage}
\begin{minipage}{.5\linewidth}
    \centering
    \includegraphics[width=.9\linewidth]{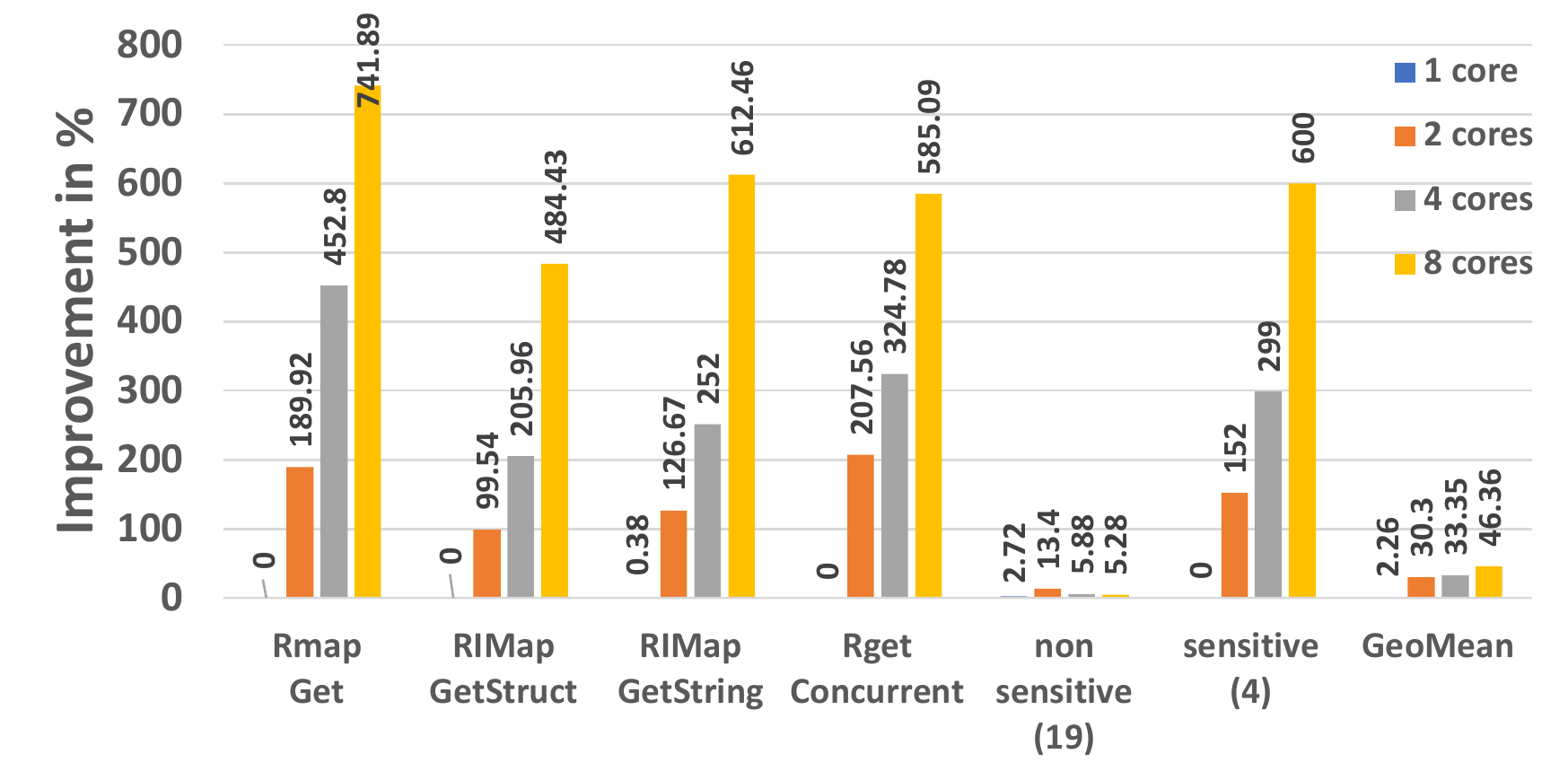}
    \caption{Results on go-cache with different core numbers. (benchmark names reflect abbreviated names of go-cache's benchmark functions).}
    \label{fig:go-cache}
    \end{minipage}
\begin{minipage}{.5\linewidth}
    \centering
    \includegraphics[width=.9\linewidth, height=4cm]{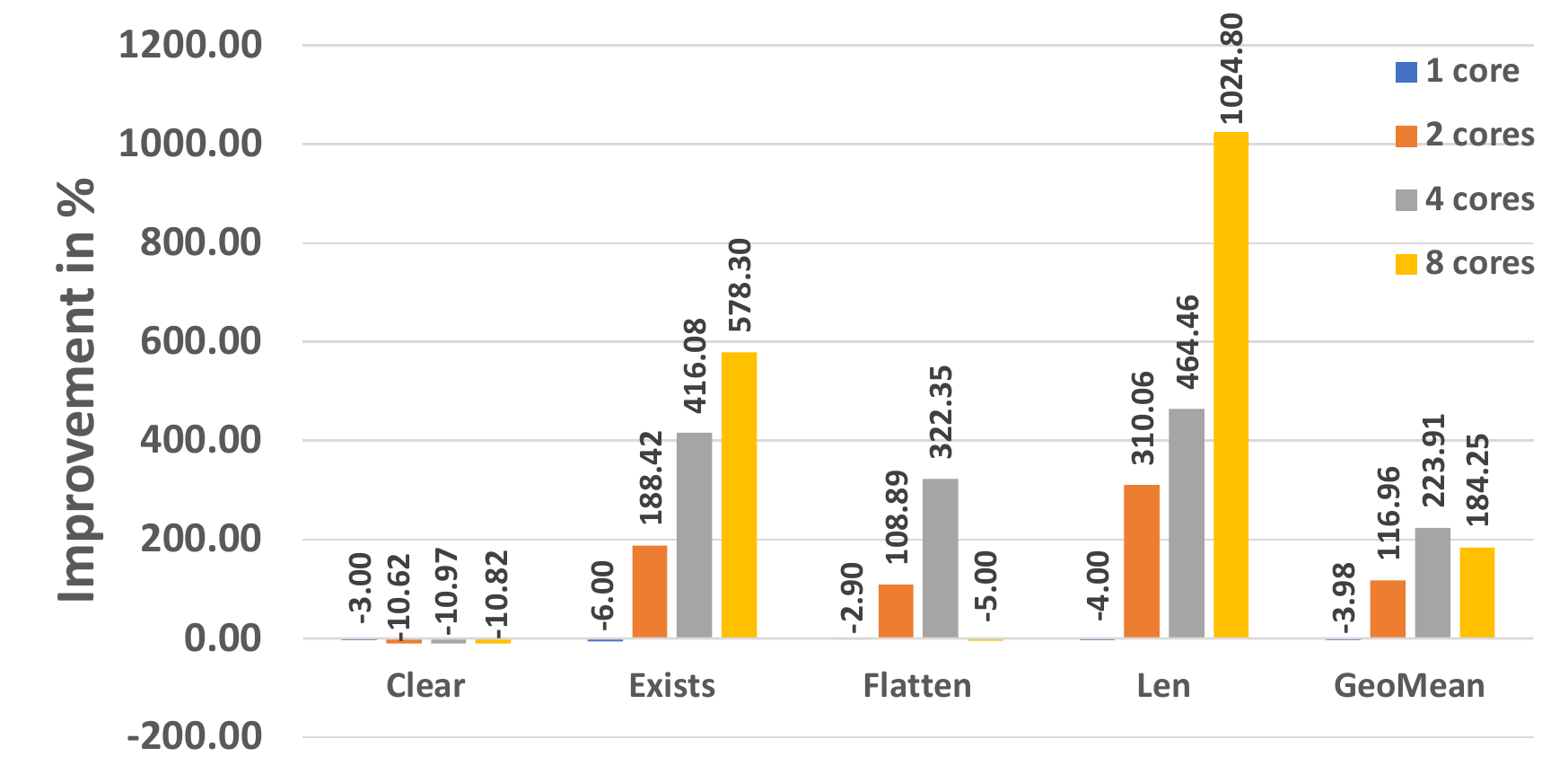}
    \caption{Results on concurrent set with different core numbers.}
    \label{fig:set}
    \end{minipage}
\begin{minipage}{.5\linewidth}
    \centering
    \includegraphics[width=0.9\linewidth]{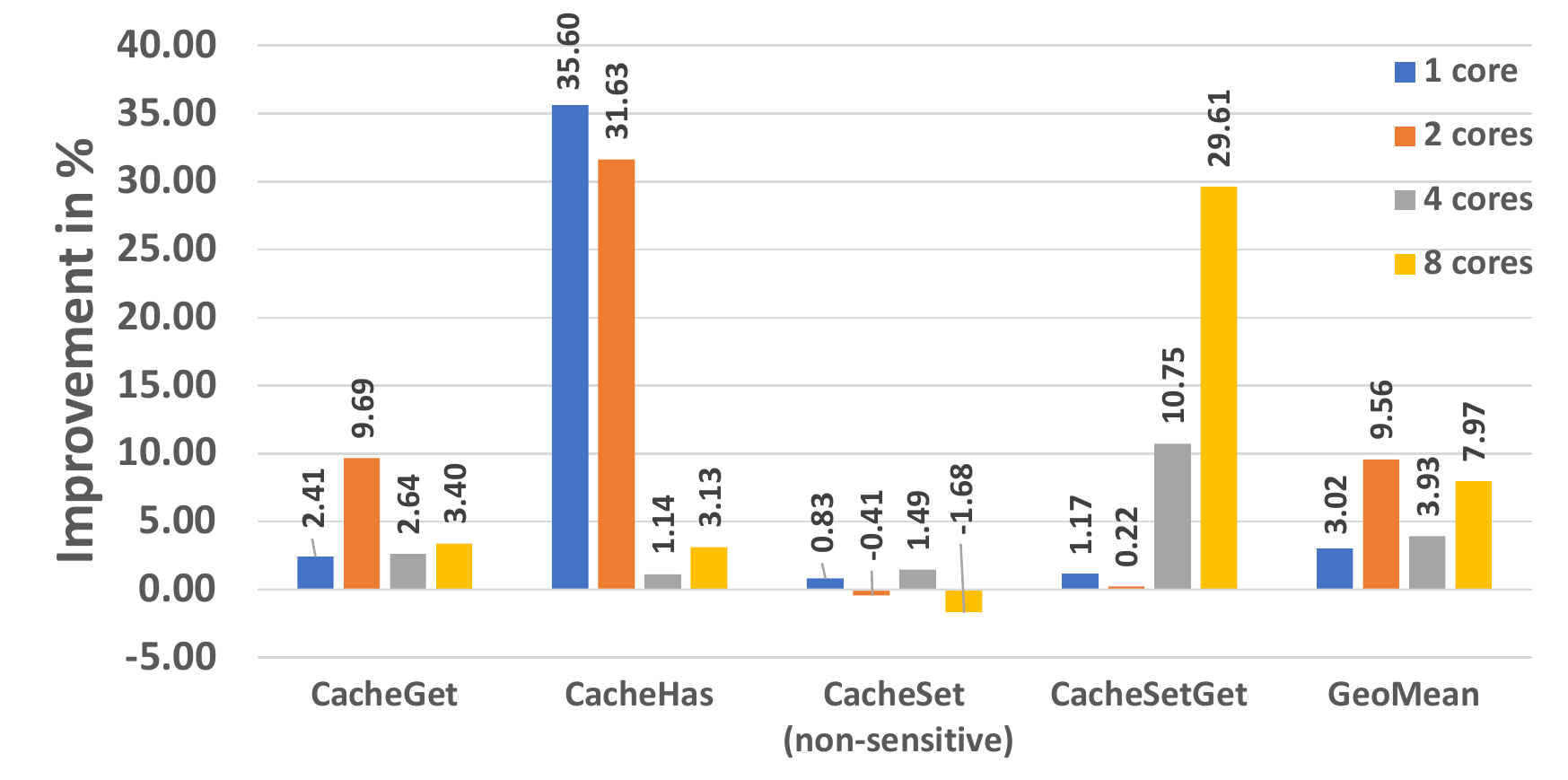}
    \caption{Results on fastcache with different core numbers.}
    \label{fig:fastcache}
\end{minipage}    
\end{figure*}
\subsection{Results on Popular Go Programs}
We categorize the benchmarks in each package into two groups:
\begin{enumerate}[noitemsep,left=0pt,topsep=1pt]
    \item \textbf{Concurrency non-sensitive} benchmarks either have no locks or do not spend much time in critical sections, or our transformation does not result in any performance difference. For these benchmarks, we only show the aggregate (geomean) results unless noted otherwise. They appear as ``non sensitive'' in our charts, and the number in the parenthesis indicates how many benchmarks are in this group.
    \item \textbf{Concurrency sensitive} benchmarks exercise modified locks  non-trivially. We might have impacted them positively or negatively. For these, we present data from each benchmark and also present an aggregate result (``sensitive'' in our charts).
\end{enumerate}
The ``all'' part of our charts is the geomean taken over all benchmarks. 
Sometimes this number looks small because of a large number of non-sensitive benchmarks.

In what follows, we provide details of performance evaluation on the aforementioned Go packages. The total number of benchmarks is large; hence, we dive deep only into benchmarks with surprisingly good speedups.

\textbf{Tally}~\cite{tally} is a fast, buffered stats collection library and Figure~\ref{fig:tally} shows its results.
For the \texttt{HistogramExisiting} benchmark, \tool{} achieves more than 660\% speedup on 8 cores reducing the original time per operation from 65 ns/ops down to around 8.47 ns/ops at 8 cores. Moreover, the HTM delivers scalable performance.
This benchmark uses a \texttt{Mutex} lock on a read-only \texttt{Exists} operation, and hence, is a natural candidate to demonstrate speedup as 
HTM eliminates the unnecessary serialization.
Conversely, the baseline has a scalability collapse, where the time per operation increases from 20.4 ns/ops to 65 ns/ops for 1 to 8 cores.
\texttt{ScopeReporting1} holds three independent \texttt{RWMutex}es at different points in time and accesses read-only data. 
However, since the \texttt{RWMutex} also involves a counter increment and a decrement, its overhead as a result of cache invalidation does not scale well.
Thus, even eliding \texttt{RWMutex} proves highly beneficial.
The speedup for \texttt{ScopeReporting10} is lower than that for \texttt{ScopeReporting1} because it performs 10x more work inside the critical section.
Overall, in the sensitive group, we see a 10\% performance drop with a single CPU but 145\%, 283\%, and 267\% improvements with 2, 4, and 8 CPUs, respectively.
In the non-sensitive group, the overall performance drop is within the margin of error.
Among all the 27 benchmarks of tally, we see up to 18.7\% speedup at 8-CPUs.

\textbf{go-cache}~\cite{patrickm61:online} is an  in-memory key-value store.
It contains benchmarks that exercise repeatedly accessing the same item in a small map.
The benchmarks contain both non-cached accesses, similar to how go programmers often use a map, and cached accesses provided by the \texttt{go-cache} layer to demonstrate the effectiveness of the library.
All benchmarks employ  \texttt{RWMutex}s for concurrent map read access.
Unlike the rest of the use cases, the benchmark files themselves contain locks, which \tool{} transforms into using HTM.

Figure~\ref{fig:go-cache} shows our empirical results.
\tool{} speeds up four benchmarks in \texttt{go-cache} that were directly accessing the map without the library-provided cache. 
In each case, we can see more than 100\% speedup; the biggest speedup is 742\%. 
The speedups come from eliminating contended atomic operations involved in entering and exiting from a reader lock.
The performance scales well with increased parallelism because while the lock-based approach incurs more and more contention, the HTM approach remains conflict-free throughout.
The other benchmarks, the majority of which employ the \texttt{go-cache}, are mildly improved, but more importantly, they were not degraded as a result of transformation via \tool{}.

\textbf{go-datastructures}~\cite{go-datastructure} is a collection of performant, thread-safe data structures. 
We apply \tool{} on the \texttt{set} subdirectory, which contains concurrency benchmarks.
The results are shown in Figure~\ref{fig:set}.
The \texttt{Len} benchmark computes the length of the set, and it is sped up by $\sim$1000\%  in the 8 cores setting.
\texttt{Len} has a short critical section that has a higher entry and exit cost due to atomic operations when using a\texttt{RWMutex}. 
HTM performance shows scalability since the HTM version remains conflict-free, whereas the lock-based version collapses with increased contention.
The \texttt{Exists} benchmark is similar to \texttt{Len}, where each goroutine searches one item in a set containing only one item.
It scales almost as well as \texttt{Len}, but more work is done in the critical section, which amortizes \texttt{RWMutex}'s overhead, and slightly reduces HTM's advantage.
The \texttt{Flatten} benchmark reads 50 elements from a shared map into a private array, with a layer of caching that eliminates repeated map scanning. It holds a \texttt{Mutex} to serialize concurrent accesses to the map/cache. The HTM version avoids the serialization and shows scalable performance for 1-4 cores.
At 8 cores, the number of conflicts resulting from updating the cache rises, which makes perceptron not use the HTM, and hence there is no speedup.
The \texttt{Clear} benchmark has true conflicts, and there is no speedup, but the HTM does not significantly degrade the performance.
Overall, utilizing \tool{} results in more than 100\% geomean performance gain while introducing less than 4\% slowdown in a single core setting.

\textbf{Zap}~\cite{zapRepo} is a library that implements fast and
structured logging in Go. 
Being a logging library, it has several IO operations, and hence \tool{} rewrote fewer locks.
Compared with other repositories, the improvement on zap is relatively mild. 
Due to arguably mild speedups on Zap, a large number of benchmarks, we omit a deeper analysis of Zap results.
Slowdowns are rare, the biggest being 7\%.
Overall, we observed a mild $\sim 4\%$ geometric mean speedup with the best case 28\% speedup.

\textbf{Fastcache}~\cite{fastcache} is a fast, scalable, in-memory cache.
The transformed code delivers a maximum of  35.60\% speedup and a geomean of 15.65\% speedup across all benchmarks.

In the \texttt{CacheGet} benchmark, goroutines repeatedly invoke the \texttt{Get} function, which uses an \texttt{RWMutex} to protect a shared map. 
\texttt{Get} has inter-procedural nested but non-conflicting locks, all of which are transformed into HTM.
\texttt{Get} looks up a key in the map and returns a value blob. 
The critical section of \texttt{Get} contains a few atomic add instructions, which update shared variables.
Transactional conflicts on the shared atomic adds are fewer at low core numbers, and the speedup is visible; however, at larger core counts, the conflicts increase, and the speedup vanishes. 
Fortunately, the perceptron kicks in and avoids any performance collapse.

The \texttt{CacheHas} benchmark is virtually the same as \texttt{CacheGet}, but its critical section is shorter since it does not return a populated value buffer.
Hence, the speedups are higher due to fewer conflicts, but it follows the same performance pattern as \texttt{CacheGet}.

In the \texttt{CacheSetGet} benchmark, each goroutine has two loops: the first loop repeatedly invokes \texttt{Set} and the second loop repeatedly invokes \texttt{Get}.
The \texttt{Set} function, which inserts a key-value pair into the map, may raise a \texttt{panic} if  certain constraints are violated. Hence, \tool{} does not modify a \texttt{Lock()} present in \texttt{Set}.
The \texttt{Get} function is already described previously.

Since all goroutines first attempt \texttt{Set}, where Go's default locks are being used, the runtime recognizes it as a starved mutex and takes away the time slice of some of the goroutines.
This runtime behavior reduces the number of lock contenders and, as a result, a few goroutines monopolize the lock. These goroutines quickly finish their series of \texttt{Set} operations and proceed into calling \texttt{Get} in a loop. The contention is lower on \texttt{Get} also since the load is now split between \texttt{Get} and \texttt{Set} with some goroutines on hold. The net effect is a high throughput for the whole benchmark. 

It is worth noting that the only other benchmark which invokes the \texttt{Set} function is the non-sensitive benchmark \texttt{CacheSet}. Even though \texttt{CacheSet} exhibits no performance improvement, and \texttt{CacheGet} shows mild performance improvement, their composition in \texttt{CacheSetGet} leads to secondary effects causing much higher performance gain at higher core counts.

\subsection{Perceptron Evaluation}\label{eval:perceptron}
\begin{figure}[!t]
    \centering
        \includegraphics[width=.9\linewidth]{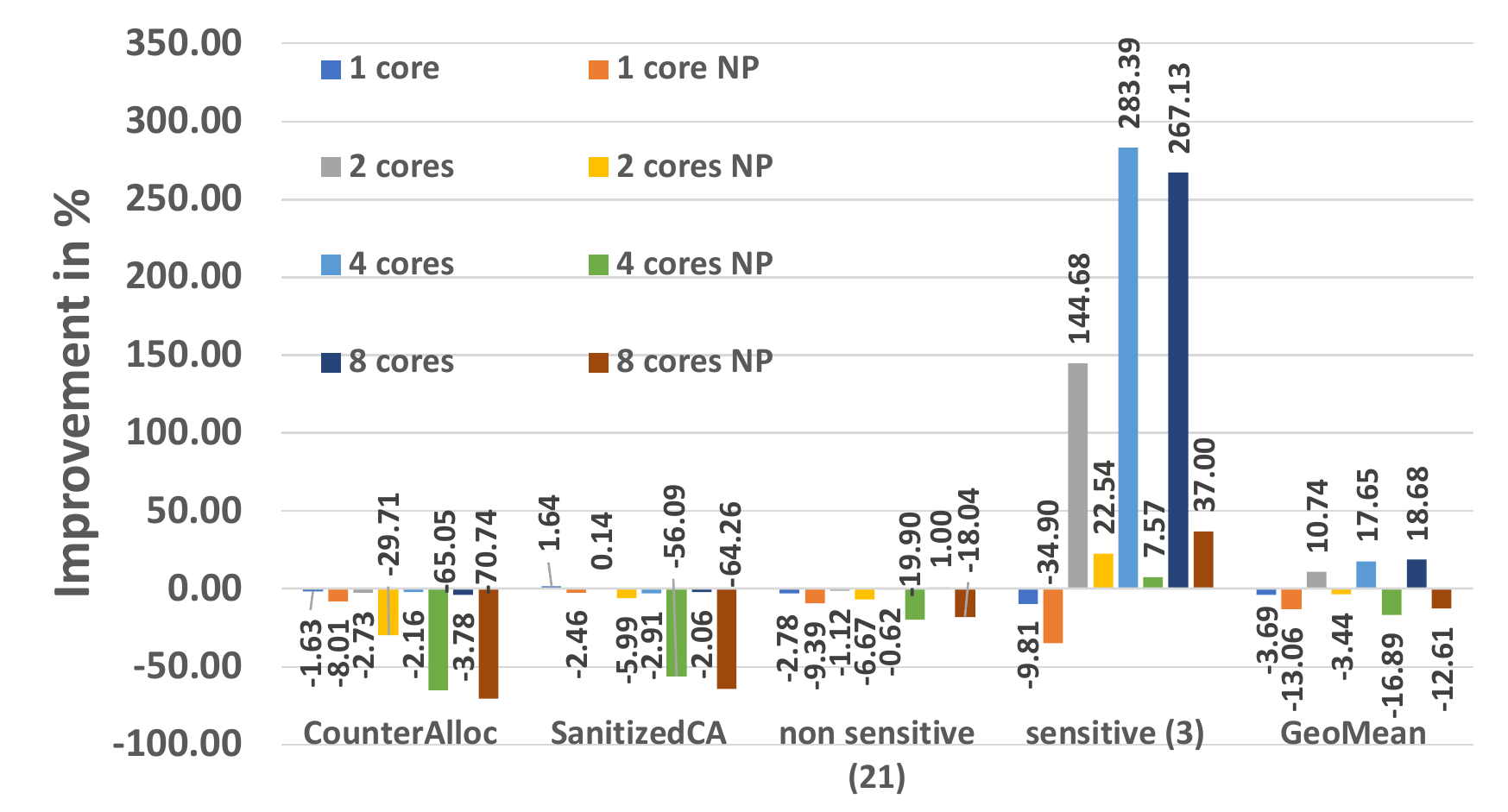}
    \caption{Results on \texttt{Tally} to show the effectiveness of perceptron. 
    NP$\implies$No Perceptron.}
    \label{fig:perceptron}
\end{figure}

We assess the effectiveness of perceptron using the \texttt{Tally} benchmarks.
We compare the performance with and without the perceptron machinery.
In the absence of the perceptron, we always attempt HTM.
In the results presented in Figure~\ref{fig:perceptron}, we can observe that the perceptron is effective in eliminating any performance loss. 
For example, CounterAllocation and SanitizedCounterAllocation are HTM-unfriendly benchmarks and cause aborts frequently.
Perceptron quickly learns to move away from HTM and keeps using the slowpath.
Therefore, there is minimal performance loss for the perceptron case.

Finally, we setup a synthetic benchmark --- a conflict-free critical section with 1000 counter updates --- to evaluate the overhead of the perceptron machinery.
We measured the perceptron prediction overhead to be 0.65\% and weight update overhead to be 0.73\% for a total of only 1.38\%.

 \section{Conclusions}
\tool{} is a source-to-source transformation tool to speed up lock-based pessimistic concurrency control in Go programs with Hardware Transactional Memory.  \tool{} combines thorough static analysis with intelligent runtime control to expose additional parallelism available in Go programs. 
 \tool{} keeps the developer in the loop, minimizes code changes via execution profiles, and targets only those critical sections that are likely to improve with HTM.
 The experimental results from real-world Go packages show that \tool{} delivers significant, scalable performance for concurrent Go code that uses locks while exhibiting rare and relatively small slowdowns.

\section{Availability}
\tool{} is available as an open-source tool~\cite{goccgithub}.

\section{Acknowledgement}
This material is based upon work supported in part by the National Science Foundation under Gran No. 1763699, 1717779, 1563935.
We thank our shepherd Michael Spear and the anonymous reviewers for their feedback.
\clearpage

\bibliographystyle{plain}
\bibliography{main}

\clearpage

\begin{appendices}
\section{Justification for using the dominance and post-dominance relationship}
\label{sec:dompdom}
The \textsc{Dom} and \textsc{PDom} requirements proposed in Definition~\ref{def:feasible}:Condition(2) may seem rather too strong.
In a more complex control flow shown in Listing~\ref{lst:complexCFG1},  Condition (2) does not hold good because neither \lp{} dominates any \up{}; however, both \lp{}s taken together guarantee a \lp{} execution before executing an \up{} and similarly, both \up{} taken together ensure that after executing any \lp{}, one \up{} is guaranteed to execute.
Thus, it would be valid to transform all \lup{} with HTM here. 

A slightly different control flow in Listing~\ref{lst:complexCFG2}, however, shows that the lack of \textsc{Dom} and \textsc{PDom} relationship is not easy to handle. In this case, the execution of the \lp{} does not ensure the execution of \up{} also or vice versa.
Hence, transforming both \lup{} does not provide any guarantee.

Any ``umbrella covering'' analysis may artificially drag non-critical sections into critical sections.
With this observation, we enforce condition (2). %
\begin{figure}[t!]
\begin{minipage}{\linewidth}
\begin{lstlisting}[language=Go,label={lst:complexCFG1},caption=Complex locking control flow amenable for HTM.]
if cond1 {
  m.Lock()
} else {
  m.Lock()
}
if cond2 {
  m.Unlock()
} else {
  m.Unlock()
}
\end{lstlisting}
\end{minipage}
\begin{minipage}{\linewidth}
\begin{lstlisting}[language=Go,label={lst:complexCFG2},caption=Complex locking control flow unsuitable for HTM.]
if cond1 {
  m.Lock()
  // in critical section
} 
// may be critical section
  
if cond2 {
  // in critical section
  m.Unlock()
} 
\end{lstlisting}
\end{minipage}
\end{figure}

\section{Splicing SESE regions for maximal LU-pairing}
\label{sec:straightlinedom}
\begin{figure}[!t]
    \centering
    \includegraphics[width=.3\linewidth]{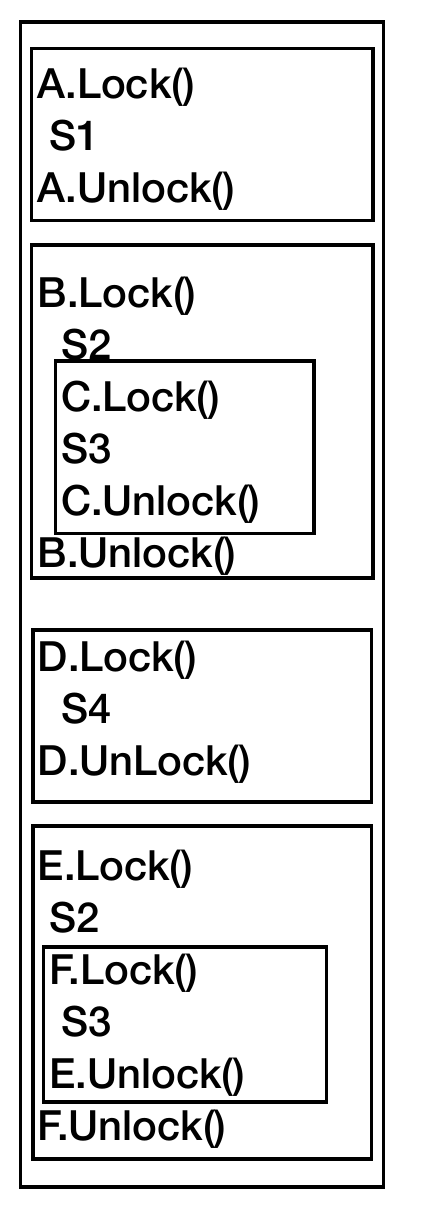}
    \caption{A straight-line sequence of statements that fall into a single region are further spliced into regions based on matching a \lp{} with the nearest post-dominating \up{} and \up{} with the nearest dominating \lp{}.}
    \label{fig:stlinedom}
\end{figure}

Although the classic definition of a program structure tree PST~\cite{JohnsonPLDI94} provides hierarchical access to SESE regions, one may miss certain lock matching opportunities if multiple lock and unlock statements happen in a straight-line sequence, as shown in Figure~\ref{fig:stlinedom}.
In this example, there is no conditional execution or loops from the first statement to the last hence, even though \lup{} split the basic blocks, 
the standard SESE would place all such basic blocks in a single innermost region without further nesting.
This would complicate which lock to pair with which unlock since the region has more than one \lupair{}.

We solve the pairing problem in such straight-line code sections by performing additional processing to splice a region further based on \lup{}.
The idea is simple; we utilize the dominator and post-dominator trees~\cite{cooperbook2007} for the region.
Only the nodes that contain a \lp{} are of interest in the dominator tree ($DomTree$).
Only the nodes that contain an  \up{} are of interest in the  post-dominator tree ($PDomTree$).

We perform a post-order traversal of the $DomTree$.
For each \lp{} $L$ in this traversal in $DomTree$, we lookup  $L$'s immediate post-dominator in the $PDomTree$, which is an \up{} $U$.
We then perform the reverse test, we lookup the immediate dominator of $U$ in the $DomTree$ that is a \lp{}, say, $\hat{L}$.
If $L = \hat{L}$, we have the innermost matching of a candidate LU-pair.
We also check that their points-to sets intersect.
If no matches are found, we traverse up in the $DomTree$ till a match is found and drop the candidate $L$ if none is found.

Matched pairs are no longer considered when matching other \lp{}s or \up{}s during the rest of the post-order traversal of nodes higher-up in the $DomTree$.
This strategy breaks a straight-line sequence into the maximum number of \lupair{}s, as shown with demarcated SESE-regions in Figure~\ref{fig:stlinedom}.

\section{Interoperability of lock-nesting with HTM}
\label{sec:locknesting}
Programs may use perfect nesting as shown in Listing~\ref{lst:perfectnest} or imperfect nesting as shown in Listing~\ref{lst:imperfectnest} (aka hand-over-hand locks~\cite{ChakrabartiAtlas2014, Kelly2020}).

\begin{figure}[t!]
\begin{minipage}{\linewidth}
\begin{lstlisting}[language=c,label={lst:perfectnest},caption={Perfectly nested locks.}]
    a.Lock()
        b.Lock()
        b.Unlock()
    a.Unlock()
\end{lstlisting}
\end{minipage}
\begin{minipage}{\linewidth}
\begin{lstlisting}[language=c,label={lst:imperfectnest},caption={imperfectly nested locks.}]
    a.Lock()
    b.Lock()
    a.Unlock()
    b.Unlock()
\end{lstlisting}
\end{minipage}
\label{fig:topsix}
\end{figure}

The following cases arise depending on the kind of lock nesting and whether or not the \lupair{} is converted into HTM.
\begin{itemize}
\item Neither the inner nor the outer lock is transformed to HTM. This is always safe because the behavior is the same as the original code in both perfect or imperfect nesting.
\item Both the inner and outer locks are transformed to HTM. 
If both inner and the outer locks use the fast path and the transaction commits successfully, it ensures mutual exclusion and atomicity of both critical sections, and hence it is safe; it also ensures that the entire transaction did not conflict with any other concurrent operation. If both inner and outer locks use the respective slow paths, the behavior is the same as the locks being untransformed (previous case) and hence obviously safe. 
These two are true whether perfect or imperfect nesting.
If one of the inner or outer HTM falls back to its slowpath, the behavior becomes the same as one of the following cases. 
\item \textbf{Perfectly nested locks}
\begin{description}
    \item[Only the inner lock is transformed to HTM.] This is safe because all instructions inside the HTM will appear to execute atomically and mutually exclusively to the external observer. While performing the inner (and the only)  transaction, if the inner lock gets acquired by another thread, the transaction will abort, and none of its state changes will be visible to the external world; the outer lock will continue to be held. The inner transaction may be retried or fall back on its slowpath. Falling back to the slowpath is the same as both inner and outer locks not being transformed.
    \item[Only the outer lock is transformed to HTM.] This is safe because the inner lock operation, including lock acquisition and release, will be done transactionally. If another thread acquires  either the inner or the outer lock while the transaction is in flight, the outer transaction will be aborted, and the inner lock acquisition (if already done) will be rolled back, ensuring mutual exclusion and atomicity of the entire region and the availability of the inner and outer locks to others. If the transaction commits, mutual exclusion and atomicity of the entire region is assured. 
\end{description}
\item \textbf{Imperfectly nested locks}
\begin{description}
    \item[Only the inner  lock is transformed to HTM.] 
    This is represented by transforming \texttt{b.Lock()} and \texttt{a.Unlock()} in Listing~\ref{lst:imperfectnest}.
    If at runtime,  fallback path is taken at \texttt{b.Lock()}, \texttt{a.Unlock()} will also use the fallback lock because our implementation recognizes mismatched mutexes from \texttt{FastLock()} to \texttt{FastUnlock()} on the same \optiLock{} object; hence, the behavior is as if we did no transformation; and hence it is functionally same as the original code.

    If at runtime the fastpath is taken at \texttt{b.Lock()}, at \texttt{a.Unlock()} we recognize a mismatched mutex \texttt{a} vs. \texttt{b} and abort the transaction. 
    The abort rolls back all changes done between \texttt{b.Lock()} till \texttt{a.Unlock()} and falls back to the slowpath. As stated before, slowpath is always correct, and hence the entire behavior is the same as the original untransformed code.
    \item [Only the outer lock is transformed to HTM.] 
    This is represented by transforming \texttt{a.Lock()} and \texttt{b.Unlock()} in Listing~\ref{lst:imperfectnest}.
    If at runtime, the fallback path is taken at \texttt{a.Lock()},  \texttt{b.Unlock()} will also use the fallback lock because our implementation recognizes mismatched mutexes from \texttt{FastLock()} to \texttt{FastUnlock()} on the same \optiLock{} object;  hence, the behavior is as if we did no transformation; and hence the behavior is functionally same as the original code.
    If at runtime, the fastpath is taken, at unlock time in \texttt{b.Unlock()} we recognize a mismatched mutex \texttt{a} vs. \texttt{b} and abort the transaction. The abort rolls back all changes done between \texttt{a.Lock()} till \texttt{b.Unlock()} including the inner \texttt{b.Lock()} and \texttt{a.Unlock()}, thus discarding all changes in the entire region. 
    Subsequently, we'll always take the slowpath route due to mismatched mutexes, and hence the effect is the same as both inner and outer locks not being transformed, which is always safe.
\end{description}
\end{itemize}

It is straightforward to extend the argument to any depth of nesting using an inductive argument that assumes that $N$-level nesting works (because $2$-level nesting works) and uses the same argument as above to prove that $N+1$-level nesting is also correct.

\section{Algorithm of optiLock}
\label{sec:optilocalgo}
Listing~\ref{lst:algo} provides a sketch of the algorithm we use in \lib{}. 

\begin{figure}[!t]
\begin{lstlisting}[language=Go,label={lst:algo},caption=Pseudo-code of FastLock() and FastUnlock()]
func (ml *OptiLock) FastLock(l *sync.Mutex) {
    ml.lkMutex = l
    compute indices and fetch perceptron weights 
    if lockCounter > threshold {
        reset  perceptron weights 
    }
    if perceptron decision == HTM {
        trial := MAX_ATTEMPTS
        for {
            if trial > 0 {
                if lock already held {
                    spin with pause till lock held
                }
                status := TxBegin()
                if status == Txstarted {
                    if lock is held {
                        abort LockHeldError
                    }
                    return
                } else {
                    trial--
                    if (aborted for LockHeldError) {
                        continue // retry
                    } 
                    if ( ! TxAbortRetry) {
                        // includes MutexMismatchError aborts
                        slowpath := true
                        htm fails := true
                        take original lock
                        break
                    }
                    // retry 
                }
            }
        }
    } else {
        lockCounter++
        slowpath = true
        take original lock
    }
}

func (ml *OptiLock) FastUnlock(l *sync.Mutex) {
    if slowPath {
        l.Unlock()
        if htm fails == true {
            compute indices decrease perceptron weights
        }
    } else {
        if l != ml.lkMutex {
            abort MutexMismatchError
        }
        TxCommit()
        compute indices and increase perceptron weights
        lockCounter = 0
    }
}
\end{lstlisting}
\end{figure}

\end{appendices}
\end{document}